\documentclass[twocolumn,twocolappendix]{aastex631}
\usepackage{CJK}
\usepackage{multirow}
\usepackage{xcolor}

\begin{document}
\begin{CJK*}{UTF8}{ipxm}

\title{Constraints on the dust size distributions in the HD 163296 disk from the difference of the apparent dust ring widths between two ALMA Bands}

\author[0000-0003-1958-6673]{Kiyoaki Doi (土井聖明)}
\affiliation{Department of Astronomical Science, School of Physical Sciences, Graduate University for Advanced Studies (SOKENDAI), 2-21-1 Osawa, Mitaka, Tokyo 181-8588, Japan}
\affiliation{National Astronomical Observatory of Japan, 2-21-1 Osawa, Mitaka, Tokyo 181-8588, Japan}
\correspondingauthor{Kiyoaki Doi}
\email{doi.kiyoaki.astro@gmail.com}
\author[0000-0003-4562-4119]{Akimasa Kataoka (片岡章雅)}

\affiliation{Department of Astronomical Science, School of Physical Sciences, Graduate University for Advanced Studies (SOKENDAI), 2-21-1 Osawa, Mitaka, Tokyo 181-8588, Japan}
\affiliation{National Astronomical Observatory of Japan, 2-21-1 Osawa, Mitaka, Tokyo 181-8588, Japan}

\begin{abstract}

The dust size in protoplanetary disks is a crucial parameter for understanding planet formation, while the observational constraints on dust size distribution have large uncertainties.
In this study, we present a new method to constrain the dust size distribution from the dust spatial distribution, utilizing the fact that larger dust grains are more spatially localized.
We analyze the ALMA Band 6 (1.25 mm) and Band 4 (2.14 mm) high-resolution images and constrain the dust size distribution in the two rings of the HD 163296 disk. 
We find that the outer ring at 100 au appears narrower at the longer wavelengths, while the inner ring at 67 au appears to have similar widths across the two wavelengths. 
We model dust rings trapped at gas pressure maxima, where the dust grains follow a power-law size distribution, and the dust grains of a specific size follow a Gaussian spatial distribution with the width depending on the grain size.
By comparing the observations with the models, we constrain the maximum dust size $a_{\mathrm{max}}$ and the exponent of the dust size distribution $p$. 
We constrain that $0.9 \ \mathrm{mm} < a_{\mathrm{max}} < 5 \ \mathrm{mm}$ and $p < 3.3$ in the inner ring, and $a_{\mathrm{max}} > 3 \times 10^1 \ \mathrm{mm}$ and $3.4 < p < 3.7$ in the outer ring. 
The larger maximum dust size in the outer ring implies a spatial dependency in dust growth, potentially influencing the formation location of the planetesimals. 
We further discuss the turbulence strength $\alpha$ derived from the constrained dust spatial distribution, assuming equilibrium between turbulent diffusion and accumulation of dust grains. 
\end{abstract}


\section{Introduction} \label{sec:intro}

The growth of dust grains in protoplanetary disks is a key process in planet formation. 
Although this process has been studied theoretically and experimentally, several challenges have been pointed out. 
One is the radial drift problem, in which dust grains drift towards the central star due to the gas friction \citep[e.g.,][]{Whipple1972, Weidenschilling1977mnras, Brauer2008, Birnstiel2009, Birnstiel2010_radialdrift, Okuzumi2012}.
Another is the dust bouncing/fragmentation problem, in which collisions between dust grains result in bouncing or fragmentation rather than growth \citep[e.g.,][]{Ormel2007,Blum2008ARAA,Wada2009,Zsom2010}.

Observations of protoplanetary disks provide clues for understanding the dust growth processes.
High-resolution observations with the Atacama Large Millimeter/submillimeter Array (ALMA) have revealed previously unseen structures in dust distributions, such as rings \citep[e.g.,][]{alma2015, Andrews2018}, spirals \citep{Perez2016, Huang2018_spiral}, and crescents \citep{vanderMarel2013, Fukagawa2013, Casassus2015}.
While the formation processes of these structures are still under discussion, gap carving by gas giants is one of the most widely accepted explanations \citep[e.g.,][]{Lin1979, Goldreich1980, Zhu2012}. 
Furthremore, there are other explanations such as hydrodynamic instabilities \citep{Ward2000, Takahashi2014, Tominaga2019}, magneto-hydrodynamic instabilities \citep{Gressel2015,Flock2015,Ueda2019}, and dust growth \citep{Okuzumi2016, Tominaga2021}.

Observational constraints on the dust size in the disks can be a key to understanding the dust growth process.
The maximum grain size provides insight into how much the dust growth proceeds/stops \citep[e.g.,][]{Testi2014review}.
The number of small grains is also important, as they contribute to the surface area of the dust grains and can affect physical and chemical properties such as chemical reactions \citep[][]{Furuya2016,Oberg2021review}, ionization degree \citep[][]{Okuzumi2009,Tsukamoto2022}, radiative cooling \citep[][]{Malygin2017}, and, as a result, (magneto-)hydrodynamic instabilities \citep[][]{Balbus_Hawley1991,Urpin1998VSI,Cui2022}.
Thus, dust size distribution can provide valuable insights into protoplanetary disks' physical and chemical processes.

However, constraining the dust size distribution is challenging.
The spectral index analysis of millimeter observations has shown different results ranging from a few hundred microns to centimeters depending on the individual studies \citep[e.g.,][]{Liu2017_TWHya_multi,Carrasco-Gonzalez2019,Huang2020,Sierra2021,Macias2021,Tsukagoshi2022,Guidi2022}.
Another approach using millimeter polarization due to self-scattering has led to estimates of the maximum grain size in the range of $\sim 100 \ \mathrm{\mu m}$ \citep[e.g.,][]{Kataoka2015,Kataoka2016a_HLTau,Kataoka2016b_HD,Stephens2014,Stephens2017,Hull2018,Sadavoy2019,Dent2019,Ohashi2019}.
Thus, there is currently no consensus on the dust size distribution.

The dust spatial distribution can be an indicator the disks' physical properties, including the dust size.
Larger dust grains are decoupled to the gas and accumulate more efficiently than smaller grains both radially and vertically.
In the equilibrium between the diffusion by gas turbulence and the accumulation of dust grains, the dust radial and vertical distributions are characterized by $\alpha/\mathrm{St}$, where $\alpha$ is the gas turbulence parameter \citep[][]{Lynden-Bell1974} and $\mathrm{St}$ is the dust-to-gas coupling parameter called Stokes number.
Some studies constrain the gas turbulence strength and/or the dust size from the dust radial \citep[][]{Dullemond2018, Rosotti2020, Sierra2019} and vertical distribution \citep[][]{Doi2021, Villenave2022, Liu2022,Pizzati2023mnrastemp}. 

In this study, we constrain the dust size distribution in the HD 163296 disk by focusing on the dust spatial distribution using the high-resolution observations in ALMA Band 6 (1.25 mm) and Band 4 (2.14 mm).
The disk has two clear dust rings at 67 au and 100 au \citep{Isella2018}.
In the dust-trapping scenario, larger grains accumulate more efficiently at pressure maximum, so dust rings should appear narrower at a longer wavelength, which traces larger grains.
We model the spectral index and the intensity profile assuming size-dependent dust trapping and put new constraints on the dust size distribution by comparing observations with the models.

The structure of the paper is as follows.
In Section \ref{sec:obs}, we explain the target object and observational data and fit the dust spatial distributions at each wavelength individually.
In Section \ref{sec:modeling}, we present a simple dust ring model and demonstrate how the dust size can be constrained using the wavelength dependence of the spatial distributions and the spectral index.
In Section \ref{sec:dust_size}, we constrain the dust size distribution by comparing the models with the observations.
In Section \ref{sec:dis}, we discuss the physical properties of the disk, such as the gas turbulence from the perspective of the dust spatial distribution.
We conclude our results in Section \ref{sec:concl}.

\section{Observations and Image Analysis} \label{sec:obs}

In this section, we perform model fitting of the HD 163296 disk in ALMA Band 6 (1.25 mm) and Band 4 (2.14 mm) independently. 
In Section \ref{sec:target}, we introduce the target and observational settings.
In Section \ref{sec:fitting}, we describe the model setup and show the fitting results in Section \ref{sec:fitting_res}.
After the physical modeling in Section \ref{sec:modeling}, we constrain the dust size distribution in Section \ref{sec:dust_size} based on the results in Section \ref{sec:obs}.

\begin{deluxetable}{ccccc}
    \tablecaption{Observational Parameters\label{tab:obs}}
    \tablewidth{0pt}
    \tablehead{
    \colhead{Band}  & \colhead{wavelength} & \colhead{beam} & \colhead{P.A.} & \colhead{rms}\\
    \colhead{}      & \colhead{[mm]}       & \colhead{[arcsec $\times$ arcsec]} & \colhead{[degree]} &\colhead{[K]}
    }
    \decimalcolnumbers
    \startdata
    Band 4 & 1.25 & 0.0650 $\times$ 0.0565 & -54.9 & 0.19 \\
    Band 6 & 2.14 & 0.0478 $\times$ 0.0383 & 81.7 & 0.26 \\
    \enddata
    \tablecomments{The observational parameters of the images used in this study.}
\end{deluxetable}

\subsection{target desctiptions and observations} \label{sec:target}

\begin{deluxetable}{lcc}
    \tablecaption{fixed parameters\label{tab:fixed}}
    \tablewidth{0pt}
    \tablehead{\colhead{quantity} & \colhead{value} & \colhead{}}
    \startdata
    disk parameter \\
    \hline
    inclination & $46.^{\circ}7$ &\\
    position angle & $133^{\circ}.3$ & \\ 
    \hline
    noise \\
    \hline
    Band 4 & 0.19 K  & \\
    Band 6 & 0.18 K & \\
    \enddata
    \tablecomments{The observational parameters of the images used in this study. The noise in Band 6 level is after smoothing to be the same spatial resolution in Band 4.  }
\end{deluxetable}

\begin{figure*}
    \begin{center}
        \includegraphics[width=14 cm]{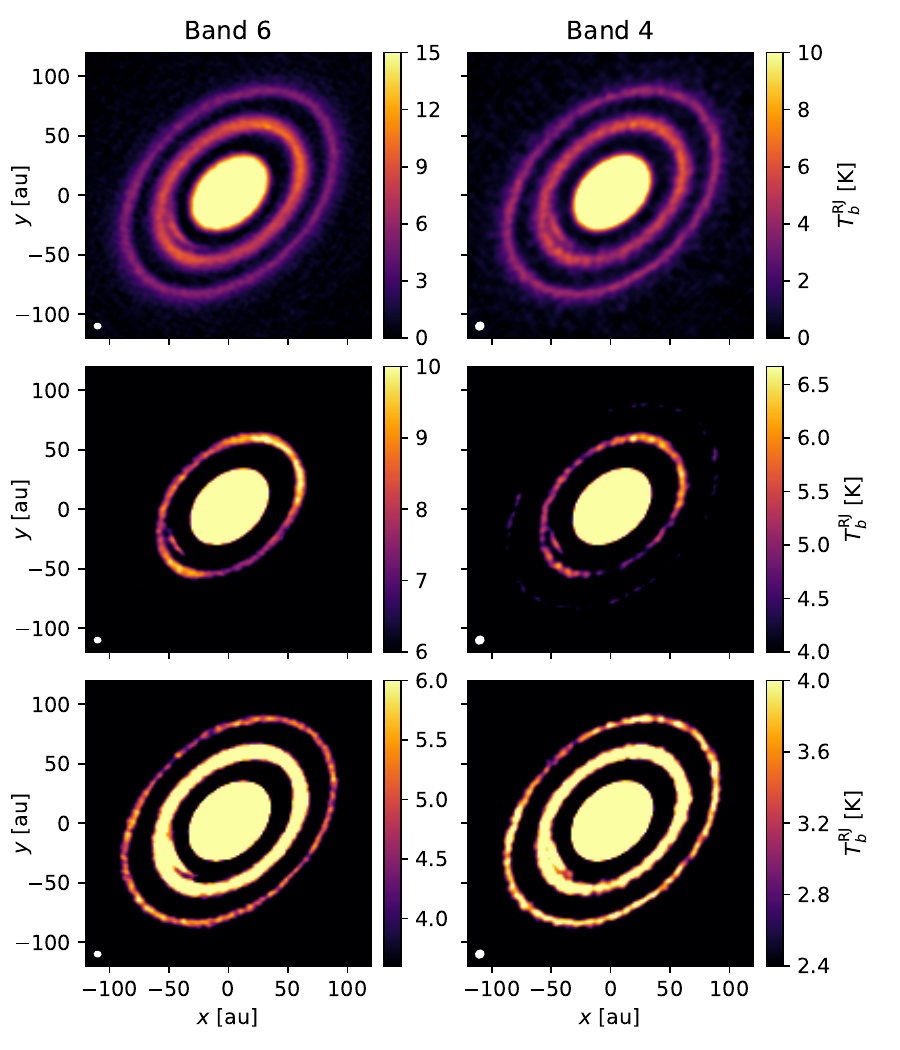}
        \caption{Observed images in Band 6 (1.25 mm) and Band 4 (2.14 mm) shown with brightness temperature assuming Rayleigh-Jeans Law.
        The middle and bottom rows show the same images as the top row but with different color ranges, which emphasize the azimuthal asymmetry of the inner ring (middle) and the outer ring (bottom).
        \label{fig:azimuthal_panels}}
    \end{center}
\end{figure*}

\begin{deluxetable*}{ccccccccc} 
        \tablecaption{The best-fit parameters. \label{tab:fitted}}
        \tablewidth{0pt}
        \tablehead{\colhead{ring} & \colhead{wavelength} & \colhead{$r_0$} & \colhead{$\tau_0$} & \colhead{$w_{\mathrm{dust}}$} & \colhead{$h_{\mathrm{dust}}$} & \colhead{$x_{\mathrm{cen}}$} & \colhead{$y_{\mathrm{cen}}$} & \colhead{$w_{\mathrm{dust,B6}}/w_{\mathrm{dust,B4}}$}\\
        \colhead{} & \colhead{[mm]} & \colhead{[au]} & \colhead{} & \colhead{[au]} & \colhead{[au]} & \colhead{[au]} & \colhead{[au]}}
        \startdata
        \multirow{2}{*}{inner}  
        & 1.25 & $67.82^{+0.05}_{-0.05}$ & $0.718^{+0.011}_{-0.011}$ & $4.82^{+0.08}_{-0.08}$ & $4.28^{+0.13}_{-0.13}$ & $0.80^{+0.06}_{-0.05}$ & $1.19^{+0.06}_{-0.06}$ & \multirow{2}{*}{$1.04^{+0.03}_{-0.03}$} \\
        & 2.14 & $67.79^{+0.07}_{-0.07}$ & $0.363^{+0.008}_{-0.007}$ & $4.65^{+0.12}_{-0.11}$ & $3.75^{+0.19}_{-0.20}$ & $1.08^{+0.07}_{-0.07}$ & $1.26^{+0.07}_{-0.08}$ \\
        \multirow{2}{*}{outer}  
        & 1.25 & $100.41^{+0.06}_{-0.06}$ & $0.423^{+0.006}_{-0.005}$ & $4.66^{+0.09}_{-0.09}$ & $< 0.57 (0.28^{+0.29}_{-0.19})$ & $0.48^{+0.08}_{-0.08}$ & $1.30^{+0.08}_{-0.08}$ & \multirow{2}{*}{$1.27^{+0.04}_{-0.04}$}  \\
        & 2.14 & $100.45^{+0.06}_{-0.06}$ & $0.322^{+0.007}_{-0.006}$ & $3.68^{+0.09}_{-0.10}$ & $< 0.97 (0.51^{+0.47}_{-0.35})$ & $0.85^{+0.09}_{-0.08}$ & $0.85^{+0.10}_{-0.10}$ \\
        \enddata
        \tablecomments{Fitted ring parameters.} 
\end{deluxetable*}

The target is the Class II protoplanetary disk around the Herbig Ae star HD 163296 with a mass of $M_{\odot} = 2.02 M_{\mathrm{sun}}$ \citep{Flaherty2015} located at a distance of $d = 101.0 \pm 0.4 \ \mathrm{pc}$ \citep{Gaia2022_DR3}.
The disk has a moderate inclination of $i = 46^{\circ}.7 \pm 0^{\circ}.1$ and a position angle of $P.A. = 133^{\circ}.1 \pm 0^{\circ}.1$ \citep[][]{Huang2018}.
Previous studies have identified millimeter dust emission consisting of a central disk and two clear rings at 67 au and 100 au (hereafter referred to as the inner and outer rings, respectively), as well as a faint ring at 155 au \citep{Isella2018}.
CO isotopologue observations \citep{Zhang2021_MAPS5} and kinematic signatures \citep{Teague2018} have revealed the presence of gas rings and gaps corresponding to the dust rings.
The disk has gas kinks, spirals, and meridional flows, which have been proposed as evidence of planets \citep{Pinte2018_planet, Izquierdo2022, Calcino2022, Teague2019_Nat}.

In this study, we analyze the high-resolution ALMA Band 4 and Band 6 images of the disk, which spatially resolve the dust rings.
The dust continuum images are shown in Figure \ref{fig:azimuthal_panels}, and the observational parameters are listed in Table \ref{tab:obs}.
We use the Band 6 image published by the DSHARP program \citep[][]{Andrews2018,Huang2018,Isella2018} and the Band 4 image published by \citet{Guidi2022}. 
The details of the data reduction and imaging of the Band 6 image are described in \citet{Andrews2018}, and those of the Band 4 image are described in \citet{Guidi2022}.
While there are other wavelength observations using VLA and ALMA \citep{Guidi2022}, we only use the Band 6 and Band 4 data because we focus on the spatially resolved observations.

\subsection{analysis procedure} \label{sec:fitting}

To make physical interpretations of the observations, we simply model the optical depth of the two rings with Gaussian distributions radially and vertically, such as
\begin{equation} \label{eq:distribution}
    d\tau(r, z) = \frac{\tau_0 \sin{i}}{\sqrt[]{2\pi} h_{\mathrm{dust}}} \exp{\left( -\frac{(r-r_0)^2}{2 w_{\mathrm{dust}}^2} - \frac{z^2}{2 h_{\mathrm{dust}}^2} \right)},
\end{equation}
where $d\tau(r,z)$ is the optical depth distribution, $\tau_0$ is the optical depth at the ring center, $i$ is the inclination of the disk, $r_0$ is the distance of the ring center, $w_{\mathrm{dust}}$ is the ring width, and $h_{\mathrm{dust}}$ is the dust scale height.
We note that the actual dust optical depth distribution may not necessarily follow the Gaussian distribution, while the model reproduces the observed intensity well, as we will discuss in Section \ref{sec:fitting_res}.
We also note that we include dust absorption and emission but ignore scattering. 
While scattering can significantly affect the estimation of the optical depth for optically thick disks, it hardly affects our results because the optical depth estimated in Section \ref{sec:fitting_res} is not optically thick.

We adopt a radially power law and vertically isothermal temperature model as
\begin{equation} \label{eq:temperature}
    T(r) = 21.8 \times \left( \frac{r}{100\ \mathrm{[au]}} \right)^{-0.5} \ \mathrm{[K]}\footnote{The temperature profile is rescaled with the update of the distance from 122 pc \citep[][]{Hipparcos1997} to 101.0 pc \citep[][]{Gaia2022_DR3}.}
\end{equation}
following \citet{Isella2016}.
In our model, $T=26.5\ \mathrm{K}$ at the inner ring and $T= 21.8\ \mathrm{K}$ at the outer ring.
There are other temperature models proposed in the literature \citep{Rosenfeld2013, Flaherty2015, Dullemond2020}, but the temperature differences at the rings are relatively small (less than 15\%).
Therefore, the choice of temperature model does not significantly affect our final results.
We investigate the case with a high-temperature limit, i.e., optically thin limit in Section \ref{app:temperature} to assess the dependence of our results on the temperature model.

\begin{figure*}
    \begin{center}
        \includegraphics[width=15 cm]{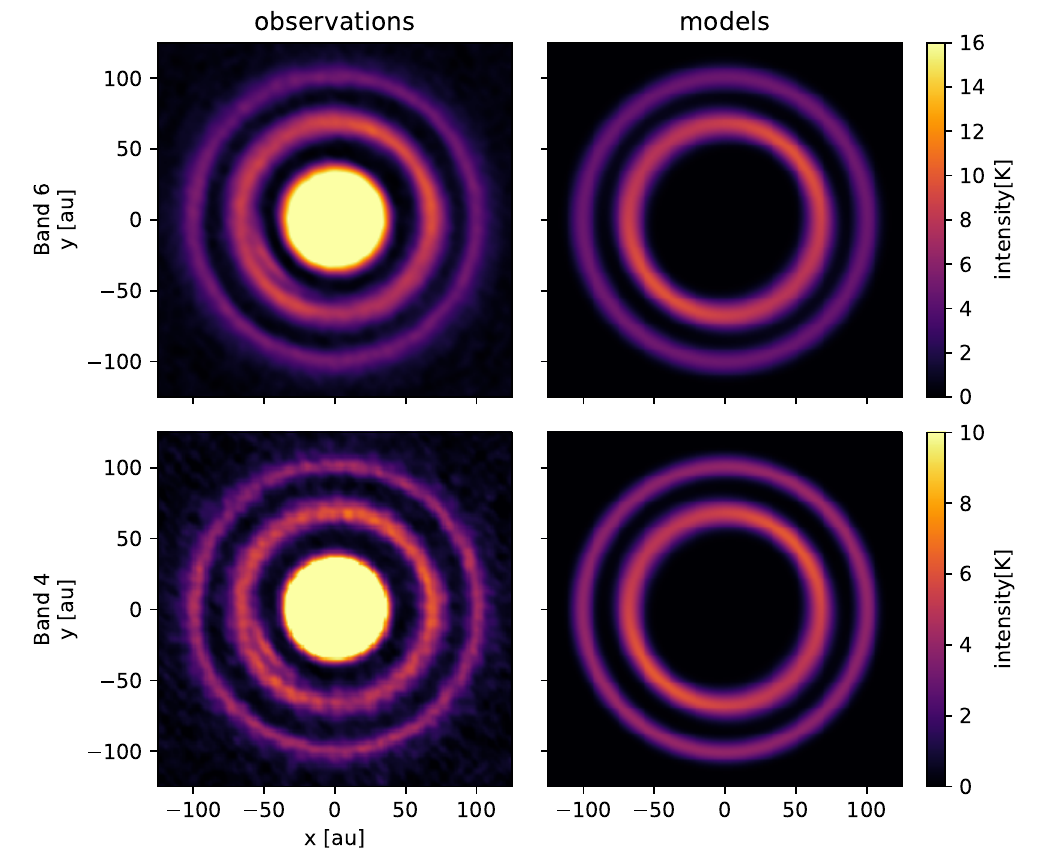}
        \caption{The deprojected images of the observations and models.
        We calculate $\chi^2$ from the residuals of these images.
        \label{fig:deproject}}
    \end{center}
\end{figure*}

We calculate the model intensity by solving the radiative transfer equation assuming the above density and temperature models.
Since we solve the radiative transfer equation along the line of sight, taking into account the disk inclination and the vertical distribution, the model intensity has a dependence on the dust scale height, especially around the disk minor axes \citep{Doi2021}.
The code used in this study is pibliclu available\footnote{\url{https://github.com/doikiyoaki/fit_ring}}.

We compare our models to the observations to constrain the following 6 parameters: $\tau_0$, $r_0$, $w_{\mathrm{dust}}$, and $h_{\mathrm{dust}}$ in equation (\ref{eq:distribution}), and the offsets of the ring center $x_{\mathrm{cen}}$ and $y_{\mathrm{cen}}$.
To avoid the impact of different angular resolution, the Band 6 observational image is smoothed with an ellipsoidal Gaussian to match the beam size of the Band 4 image.
 The model images are also smoothed with the same beam size as the observations. 
Both the model and observational images are deprojected using the inclination of $i=46.^{\circ}7$ and the position angle of $P.A. = 133.^{\circ}3$ of the disk following \citet{Huang2018}.
Figure \ref{fig:deproject} shows the deprojected images of the observations and the models. 
The noise level from emission-free regions is 0.28 K in Band 6, which reduces to 0.17 K after smoothing, and 0.19 K in Band 4. 
We treat the data points at every geometric mean of FWHM as independent data points and calculate $\chi^2$ from the residual between observations and the models around the ring ($60 \ \mathrm{au} < r < 75 \ \mathrm{au} $ for the inner ring and $94 \ \mathrm{au} < r < 106 \ \mathrm{au} $ for the outer ring).
We exclude data in the southeast direction ($180^\circ$ to $270^\circ$) to remove the contribution of the asymmetric, crescent-shaped structure \citep[][]{Isella2018}.
Since the choice of the fitting region can affect the fitting results, we also perform the fitting with the different fitting regions and confirm that our results are not affected by the choice of the fitting region (see Appendix \ref{app:comparing}).

We use the Markov Chain Monte Carlo (MCMC) method to sample the probability distribution and estimate the parameters with uncertainties.
The log-likelihood function is defined as $-0.5 \chi^2$.
We set uniform prior distributions for all parameters.
For the dust scale height and optical depth, we set the priors to be $>0$.
We use the \texttt{emcee} package \citep{Foreman-Mackey2013} for MCMC sampling.
We set 32 walkers for 10,000 steps and discard the initial 1,000 steps as burn-in.
We adopt the 50th percentile of the marginalized distribution as the estimated value and the 16th and 84th percentiles as the $1\sigma$ error range.
We also estimate the maximum likelihood estimator, which is the parameter set that maximizes the probability.
We note that for the dust scale height, the maximum likelihood estimator can approach zero, especially in the outer ring, causing it to fall outside the $1\sigma$ error range. In such case, we take the 84th percentile as an upper limit. The corner plots showcasing these instances can be found in Appendix \ref{app:mcmc}.
\\

\subsection{Fitting Results} \label{sec:fitting_res}

\begin{figure*}
    \begin{center}
        \includegraphics[width=16 cm]{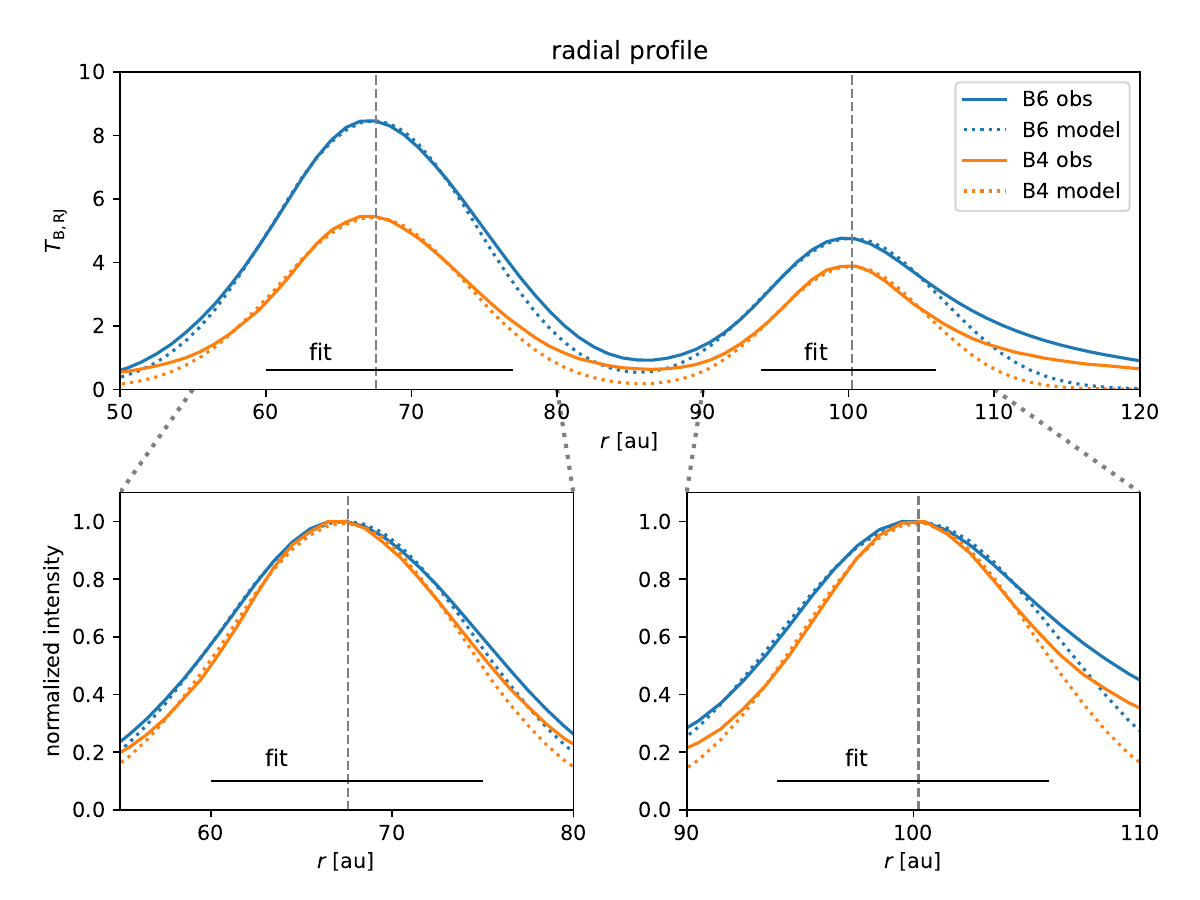}
        \caption{Azimuthally averaged radial profiles.
        Top: the observed profiles (solid lines) and the fitted model profiles (dotted lines) in Band 6 and 4.
        Bottom: the observed profiles normalized at the peak of each ring.
        \label{fig:radial_2band}}
    \end{center}
\end{figure*}

\begin{figure*}
    \begin{center}
        \includegraphics[width=16 cm]{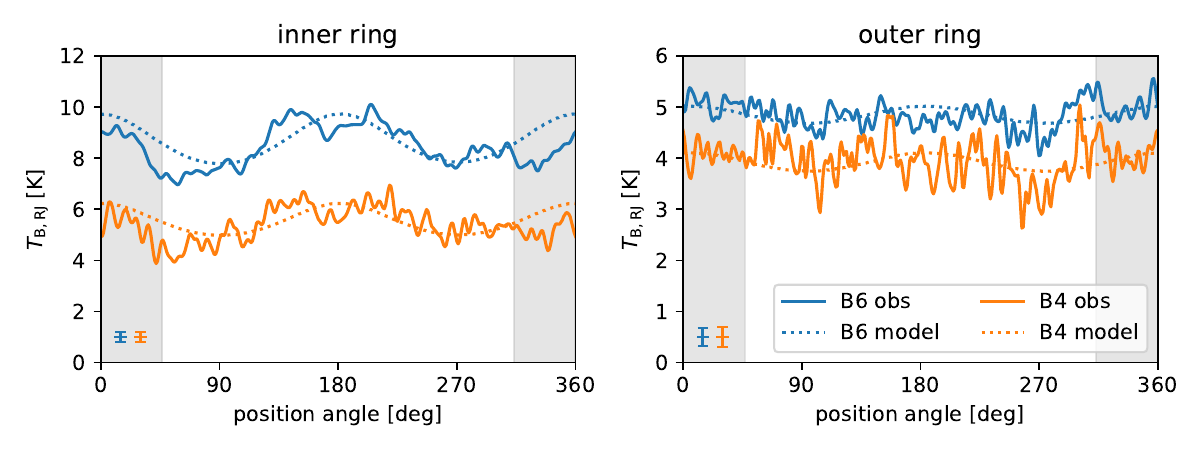}
        \caption{ Azimuthal profile of the observations and best-fit models along the two rings.
        The position angle is measured in a counterclockwise direction from the southeast major axis.
        Vertical lines at the bottom left indicate the thermal noise.
        Gray areas denote regions excluded from fitting due to the crescent-shaped non-axisymmetric structure. 
        \label{fig:azimuth_2band}}
    \end{center}
\end{figure*}

\begin{figure}
    \begin{center}
        \includegraphics[width=8 cm]{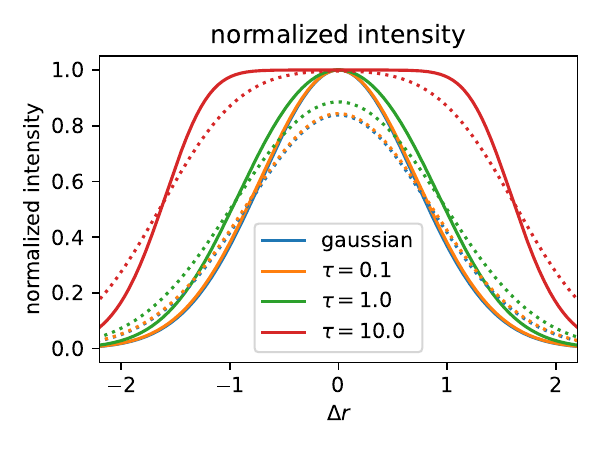}
        \caption{
        Normalized radial intensity profiles assuming Gaussian optical depth distributions.
        The width of Gaussian is $\sigma = 1$ for all models.
        The solid lines are the profiles without beam convolution, and the dotted lines are the profiles with beam convolution corresponding to the Band 6 beam size for the inner ring width ($\sigma = 2.5\ \mathrm{au}/ 4.8\ \mathrm{au} = 0.46$).
        The profile is top-flat if the ring is optically thick and the beam resolves the ring.
        \label{fig:radial_gauss}}
    \end{center}
\end{figure}

The best-fit parameters are shown in Table \ref{tab:fitted} with $1\sigma$ error range.
We show the maximum likelihood parameters in Appendix \ref{app:mcmc}.
Figure \ref{fig:radial_2band} shows the radial profiles of the observations and the model excluding the southeast direction, which has a non-axisymmetric structure.
The model and observations appear to be in good agreement within the fitting region.
However, the observed intensity exceeds the model intensity outside the fitting region. 
This may be due to the presence of small dust grains, which have extended distribution.
Also, the gas profile may deviate from the Gaussian profile, especially at the edge of the ring. 
In particular, the discrepancy between the observations and the models is large at the outside of the outer ring ($r>105 \ \mathrm{au}$). 
One possible explanation is that the pressure gradient is smaller due to the absence of planets outside of the outer ring.
Another explanation is that the dust spatial distribution has not reached an equilibrium between drift and diffusion due to the longer timescales of dust growth and advection in the outer region.

The ring width of the inner ring is fitted to $w_{\mathrm{dust,B6}} = 4.82^{+0.08}_{-0.08} \ \mathrm{au}$ in Band 6 and $w_{\mathrm{dust,B4}} = 4.65^{+0.12}_{-0.11} \ \mathrm{au}$ in Band 4, and that of the outer ring is fitted to $w_{\mathrm{B6}} = 4.66^{+0.09}_{-0.09} \ \mathrm{au}$ in Band 6 and $w_{\mathrm{B4}} = 3.68^{+0.09}_{-0.10} \ \mathrm{au}$ in Band 4. 
The ratio of the ring width in Band 6 to Band 4 is $w_{\mathrm{dust,B6}}/w_{\mathrm{dust,B4}} = 1.04^{+0.03}_{-0.03}$ in the inner ring and $w_{\mathrm{dust,B6}}/w_{\mathrm{dust,B4}} = 1.27^{+0.04}_{-0.04}$ in the outer ring. 
The dust ring width is narrower in Band 4 than in Band 6 in the outer ring, while the dust ring width is similar or slightly narrower in Band 4 than in Band 6 in the inner ring.
We confirm the trends in Figure \ref{fig:radial_2band}.
This difference in ring width at different wavelengths is expected based on dust trapping models, and we observationally confirm it.

The dust scale height in the inner ring is fitted to $h_{\mathrm{dust,B6}} = 4.28^{+0.13}_{-0.13}\ \mathrm{au}$ in Band 6 and $h_{\mathrm{dust,B4}} = 3.75^{+0.19}_{-0.20}\ \mathrm{au}$ in Band 4, and that in the outer ring is fitted to  $h_{\mathrm{dust,B6}} < 0.57\ \mathrm{au}$ in Band 6 and $h_{\mathrm{dust,B4}} < 0.97\ \mathrm{au}$ in Band 4. 
It means that the dust grains are puffed up in the inner ring and settled in the outer ring.
These trends can be confirmed visually from the azimuthal symmetry/asymmetry of the observed image in Figure \ref{fig:azimuthal_panels} and the azimuthal profile along the ring presented in Figure \ref{fig:azimuth_2band}, as discussed by \citet{Doi2021}.
The intensity is larger at the major axes than at the minor axes along the inner ring, which means that the dust grains are puffed up.
In contrast, the intensity is axisymmetric along the outer ring, which means that the dust grains are settled.
The fitted dust scale height shows similar trends as \citet{Doi2021}, with slightly different values.
In \citet{Doi2021}, the dust radial distribution is fixed and given by hand, which can affect the estimation of the dust scale height.
Here, we fit the dust radial profile and dust scale height simultaneously, which is more appropriate with fewer assumptions.

The fitted optical depth shows that the disk is optically thin at all rings and bands, with $\tau<1$.
While scattering induced intensity reduction could potentially lead to an underestimation of the optical depth \citep{Zhu2019,Liu2019scattering,Ueda2020}, we continue to assume that the rings are not optically thick for the following reasons.
First, the intensity along the inner ring exhibits azimuthal variation \citep{Doi2021}, which can be explained by changes in the optical depth along the line of sight, suggesting that the ring is not optically thick. 
Second, the radial profiles of both rings do not show a top-flat shape. 
Figure \ref{fig:radial_gauss} shows intensity distributions of models assuming Gaussian optical depth distributions. 
 A top-flat profile would be expected for an optically thick ring ($\tau=10$ model), even after beam convolution, if the ring is resolved by the beam.
The model also demonstrates that a higher optical depth results in a broader intensity distribution, even when the optical depth distributions are the same.
Therefore, we need ray-tracing simulations to estimate the ring width based on observations.

\begin{figure*}
    \begin{center}
        \includegraphics[width=15 cm]{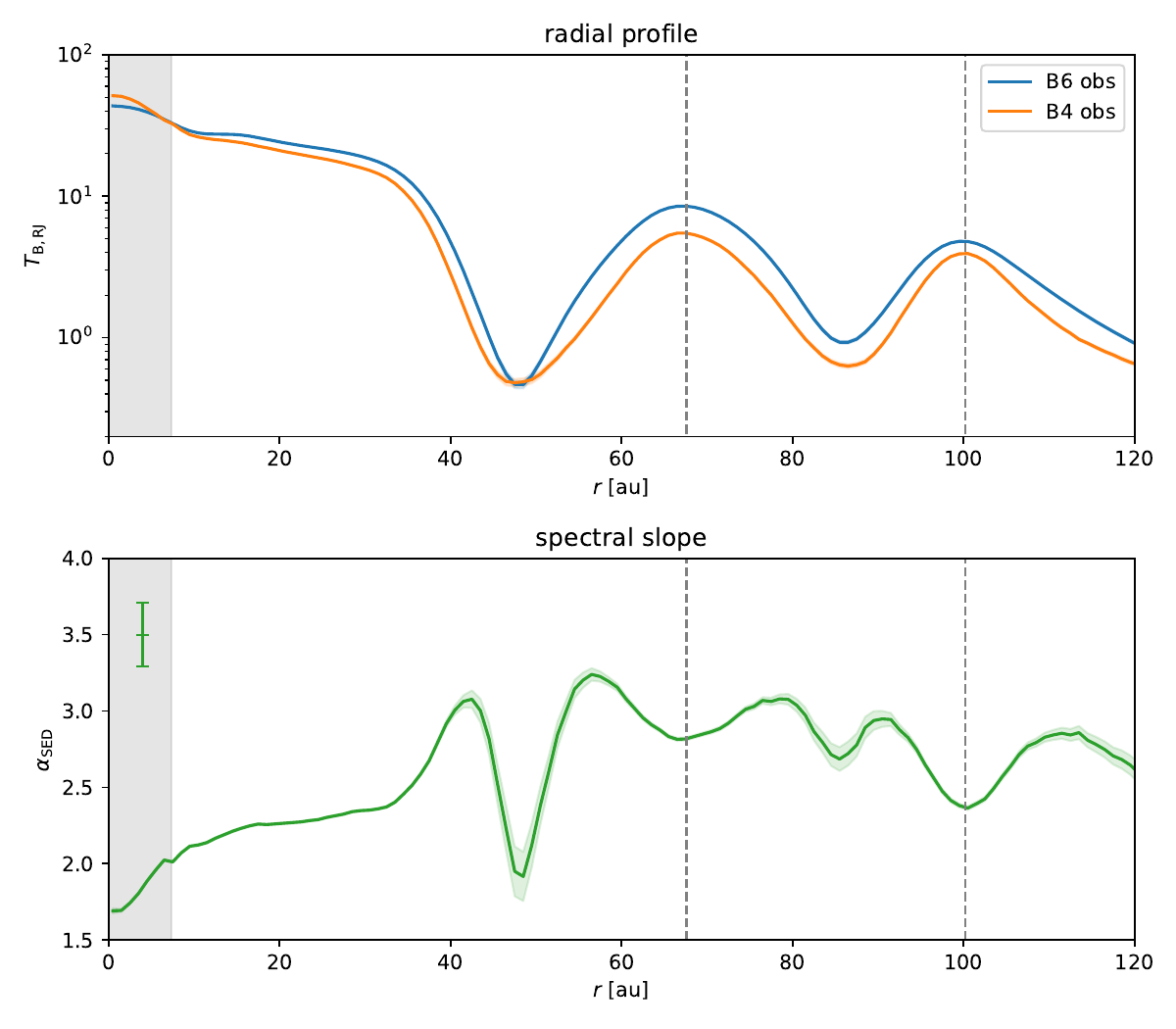}
        \caption{Top: the observed radial intensity profile in Band 6 and 4. 
        Bottom: the slope $\alpha$ of the spectral energy distribution (SED) at each radius.
        The left gray regions represent the observed beam size, and the left vertical bar represents uncertainties from the absolute flux calibration error with $1\sigma$ ranges.
        \label{fig:SED}}
    \end{center}
\end{figure*}

\section{Theoretical Modeling of Dust Ring} \label{sec:modeling}

In this Section, we construct theoretical models of the dust rings, assuming size-dependent dust trapping, to compare with the observations in Section 2 and constrain the dust size distributions.
We assume pressure trapping for both of the rings of the HD 163296 disk.
This assumption is supported by the kinematic detection of gas pressure bumps corresponding to the two dust rings \citep{Teague2018, Rosotti2020}.
Under this assumption, larger dust grains accumulate more efficiently at the pressure maximum, resulting in the dust ring width appearing narrower at longer wavelengths.

The ratio of apparent dust ring widths between two wavelengths, $w_{\mathrm{dust,\lambda_1}}/w_{\mathrm{dust,\lambda_2}}$, can serve as an indicator of the dust size distribution.
Though the dust ring width $w_{\mathrm{dust, \lambda}}$ relies on multiple factors including the gas ring width, the gas turbulence strength, the gas surface density, and the dust size distribution, we note that these gas properties remain the same within the same ring, regardress of the observed wavelength. 
Consequently, these gas-related terms are canceled out when the ratio is taken, and this ratio depends solely on the dust size distribution.

In Section \ref{sec:model_profile}, we describe the underlying assumptions on dust size distribution, spatial distribution, and opacity.
Then, we model the ring profiles in Band 6 and Band 4 to reveal the dependence of the ring width ratio on the dust size distribution in Section \ref{sec:model_width_ratio}.
In Section \ref{sec:model_sed}, we show the dependence of the opacity spectral index on dust size distribution\footnote{We use $\alpha$ to denote the slope of the spectral energy distribution, also referred to as the spectral slope, and $\beta$ to denote the spectral index of the dust opacity, similarly referred to as the opacity spectral index.}.

\subsection{Model assumptions} \label{sec:model_profile}

We describe the assumptions on dust size distribution, spatial distribution, and opacity.
We assume the number density of the dust grains across an entire ring as
\begin{equation} \label{eq:dust_size_distribution}
N(a) \propto a^{-p} \ \mathrm{for} \ a_{\mathrm{min}} \leq a \leq a_{\mathrm{max}},
\end{equation}
where $N(a)$ is the number density of dust grains with size $a$, $a_{\mathrm{min}}$ and $a_{\mathrm{max}}$ are the minimum and maximum dust sizes, and $p$ is the exponent of the power law dust size distribution.
\citet{Mathis1977} found that the exponent $p=3.5$ in the interstellar medium from infrared extinction, and similar values are often assumed in protoplanetary disks.
 Some studies derive the power law exponent from millimeter multiwavelength observations, which may yield comparable values \citep[e.g.,][]{Macias2021,Guidi2022}.
In this study, we fix the minimum dust size to be $1 \ \mathrm{\mu m}$ and constrain the maximum dust size $a_{\mathrm{max}}$ and the exponent $p$.

\begin{figure}
    \begin{center}
        \includegraphics[width=8 cm]{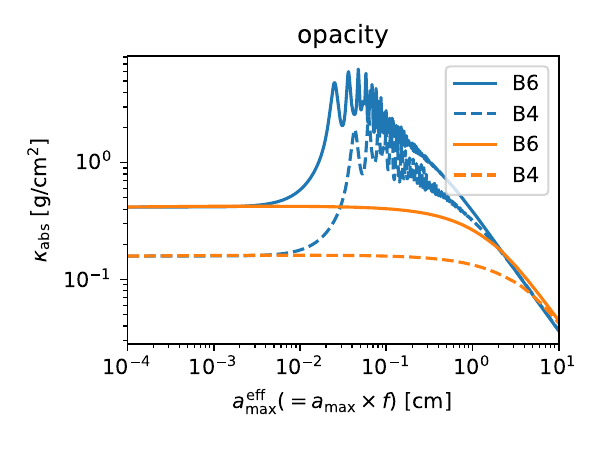}
        \caption{The dust opacity profile of compact (solid line) and porous (dashed line) dust.
        We used the dust model by \citet{Birnstiel2018}
        \label{fig:opacity}}
    \end{center}
\end{figure}

We adopt the DSHARP dust opacity model from \citet{Birnstiel2018}.
We calculate the opacity using the Mie theory with the \texttt{dsharp\_opac} package \citep{Birnstiel2018}.
Figure \ref{fig:opacity} shows the opacity of compact and porous dust grains with filling factors of $f=1$ and $0.01$, respectively.
The horizontal axis represents the effective dust size $a^{\mathrm{eff}} \equiv a \times f$, scaled with porosity \citep[][]{Kataoka2014}.
The dust opacity has a peak when the dust size is similar to the observed wavelength if the dust grains are compact, while the opacity does not have a peak if the dust grains are porous.
In the case of $\lambda > a$, the dust opacity is smaller for larger grains.
In the following discussion, we assume compact grains.
We additionally discuss cases of porous grains and different dust compositions in Appendix \ref{sec:porous}.

For the dust spatial distribution, we assume that the dust particles are trapped at a Gaussian gas pressure bump.
\citet{Dullemond2018} derived the relation between the dust ring width of single-sized dust grains $w_{\mathrm{dust}}$ and gas ring width $w_{\mathrm{gas}}$, assuming equilibrium between the diffusion of the dust by gas turbulence $\alpha$ and dust radial accumulation as
\begin{equation} \label{eq:radial-1}
w_{\mathrm{dust}} = \left( 1 + \frac{\mathrm{St}}{\alpha_r} \right)^{-1/2} w_{\mathrm{gas}},
\end{equation}
where $\alpha$ is the gas turbulence parameter and $\mathrm{St}$ is the dust-to-gas coupling parameter called the Stokes number $\mathrm{St} (= \pi \rho_{\mathrm{mat}} a / 2 \Sigma_{\mathrm{gas}})$.
Here, $\rho_{\mathrm{mat}}$ is the dust material density, $a$ is the dust radius, and $\Sigma_{\mathrm{gas}}$ is the gas surface density.
We denote the direction of the turbulence with subscript $r,z$ for the radial and vertical direction since the gas turbulence is not necessarily isotropic.
If $\alpha/\mathrm{St} \ll 1$, the equation can be approximated as
\begin{equation} \label{eq:radial_approx}
w_{\mathrm{dust}} = \sqrt[]{\frac{\alpha_r}{\mathrm{St}}} w_{\mathrm{gas}}.
\end{equation}
Since the Stokes number is proportional to the dust size, we can treat it as a dimensionless parameter of the dust size.
Therefore, the dust ring width is suggested to have the following dependency on the dust size, given by 
\begin{equation} \label{eq:radial_approx-2}
w_{\mathrm{dust}}(a) \propto 1/\sqrt[]{\mathrm{St}} \propto 1/\sqrt[]{a}.
\end{equation}
Based on these dust size distributions, spatial distributions, and dust opacity, we will further model the opacity distribution and the opacity spectral index.

\subsection{Radial profile and ring width ratio} \label{sec:model_width_ratio}

We model the optical depth distribution at two wavelengths to reveal the dependency of the dust ring width ratio on the dust size distribution.
We calculate the radial profile of the model optical depth as
\begin{equation} \label{eq:opacity_distribution}
\tau(\Delta r) = \int \frac{\kappa (a) m(a) N(a)}{\sqrt{2 \pi}  w_{\mathrm{dust}}(a)} \exp{ \left( - \frac{\Delta r^2}{2 w_{\mathrm{dust}}(a)^2} \right)} da,
\end{equation}
where $\Delta r$ is the distance from the ring center, $\kappa (a)$ is the dust absorption opacity, $m(a)$ is the dust grain mass, 
and $N (a)$ is the dust number density in the entire ring.
Here, the grain mass is given by
\begin{equation}
    m(a) = \frac{4}{3} \pi \rho_{\mathrm{mat}} a^3.
\end{equation}
We calculate the optical depth profile assuming $w_{\mathrm{dust}}(a) \propto 1/\sqrt{a}$ as equation (\ref{eq:radial_approx-2}).
Here, we normalize the radial distance by the width of the $1 \ \mathrm{mm}$ dust grain, and the dust ring width is given by
\begin{equation}
    \frac{w_{\mathrm{dust}}(a)}{w_{\mathrm{dust}} (1 \ \mathrm{[mm]})} = \sqrt{\frac{1\ \mathrm{mm}}{a}}
\end{equation}
We compute the optical depth distribution by numerically solving equation (\ref{eq:opacity_distribution}) with the dust size distribution in equation (\ref{eq:dust_size_distribution}) and the dust opacity in Figure \ref{fig:opacity}.
The equation (\ref{eq:radial_approx-2}) is not valid for the dust grains at the smallest size limit because $\alpha/\mathrm{St} \gg 1$ is not satisfied.
However, the approximation does not affect the results since the opacity of these small dust grains does not significantly contribute to the optical depth if $p<4$. 

\begin{figure}
    \begin{center}
        \includegraphics[width=8 cm]{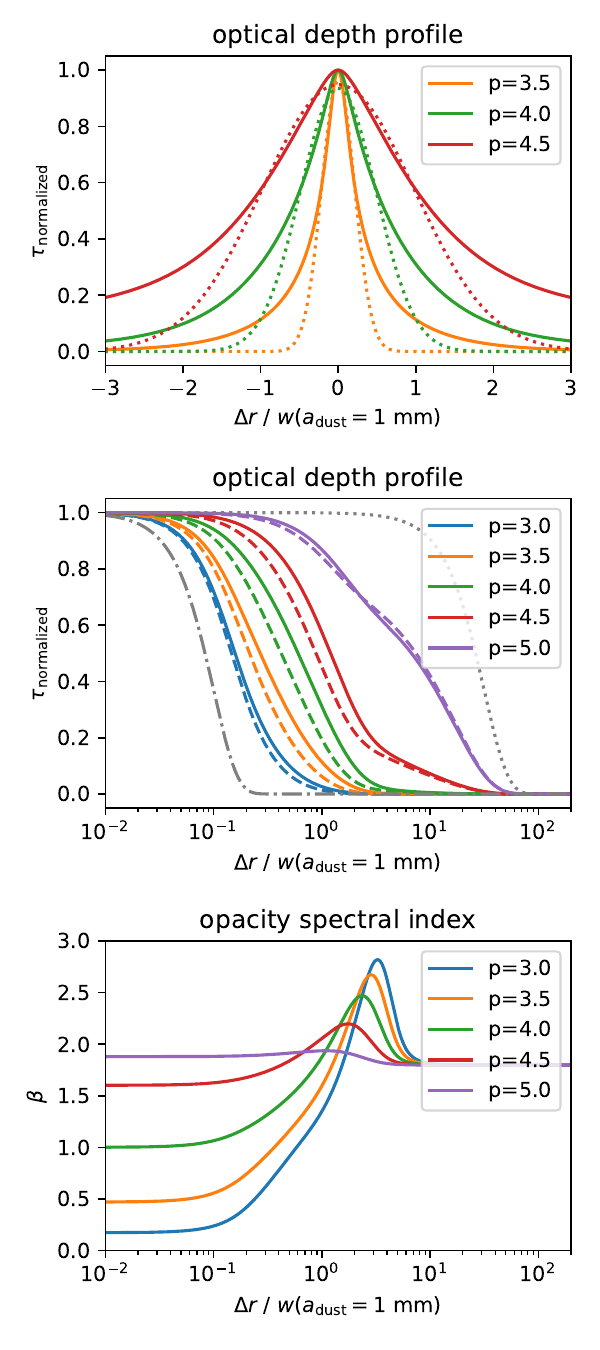}   
        \caption{
        Top: Normalized optical depth profiles in Band 6 with a linear scale.
        The horizontal axis is the radial distance normalized by the dust ring width of 1 mm grains $w_{\mathrm{dust}}(1 \ \mathrm{mm})$.
        The dotted lines are Gaussian fittings for the model profiles, and we define traced dust size in Section \ref{sec:turbulence} from these fitted widths.
        Middle: Normalized optical depth profiles in Band 6 (1.25 mm) and Band 4 (2.14 mm) with a log scale. 
        The solid lines show the opacity distributions in Band 6, and the dashed lines show those in Band 4.
        Each color corresponds to each exponent of the dust size distribution $N(a) \propto a^{-p}$.
        The gray dotted lines show the distribution of the maximum ($a=10\ \mathrm{cm}$) and the minimum ($a=1\ \mathrm{\mu m}$) dust.
        Bottom: Radial opacity spectral index profile between Band 6 and Band 4.\\
        \label{fig:model_radial}}
    \end{center}
\end{figure}

First, we present the radial optical depth profile of the model. 
The top panel of Figure \ref{fig:model_radial} shows the normalized model profile for $p = 3.5, \ 4.0,\ 4.5$, assuming a maximum dust size $a_{\mathrm{max}} = 10 \ \mathrm{cm}$, which is much larger than the observed wavelengths. 
The dotted lines represent the Gaussian function fitted to the regions with $\tau_{\mathrm{normalize}} > 0.5$. 
The fitted profiles match well with the models around the ring center, but the optical depth is larger at the rings' edge because the model profile is a superposition of Gaussian functions with different widths.
The larger optical depth than the Gaussian function at the edge is consistent with the observations.

Then, we investigate how the exponent $p$ affects the difference in the optical depth profile between Band 6 (1.25 mm) and Band 4 (2.14 mm).
The middle panel of Figure \ref{fig:model_radial} shows the optical depth profile in Band 4 and Band 6 for various values of $p$ ranging from 3 to 5, normalized by the optical depth at the ring center assuming the maximum dust size $a_{\mathrm{max}} = 10 \ \mathrm{cm}$.
When $p = 3.5,\ 4.0,\ \mathrm{and} \ 4.5$, the dust ring appears narrower in Band 4 than in Band 6 because the longer wavelength traces larger dust grains. 
However, there is no difference between the two wavelengths when $p= 3.0 \ \mathrm{or} \ 5.0$ because the dust opacity is dominated by the smallest or largest dust grains, respectively.

The bottom panel of Figure \ref{fig:model_radial} shows the radial profile of the spectral index of the dust opacity defined as
\begin{equation} \label{eq:def_beta}
    \beta = \frac{\log{\tau_{\mathrm{B6}}/\tau_{\mathrm{B4}}}}{\log{{\nu_{\mathrm{B6}}}/\nu_{\mathrm{B4}}}}
\end{equation}
The opacity spectral index $\beta$ reaches a minimum at the ring center and a maximum where the dust grains with peak opacity dominate the opacity. 
If $p=5.0$, the opacity spectral index is almost constant across the ring because the minimum dust size dominates the opacity, even at the ring center. 
The opacity spectral index at the ring center depends on the exponent $p$, with smaller $p$ values resulting in smaller $\beta$ at the ring center if the maximum dust size is significantly larger than the observed wavelength.

\begin{figure*}
    \begin{center}
        \includegraphics[width=17 cm]{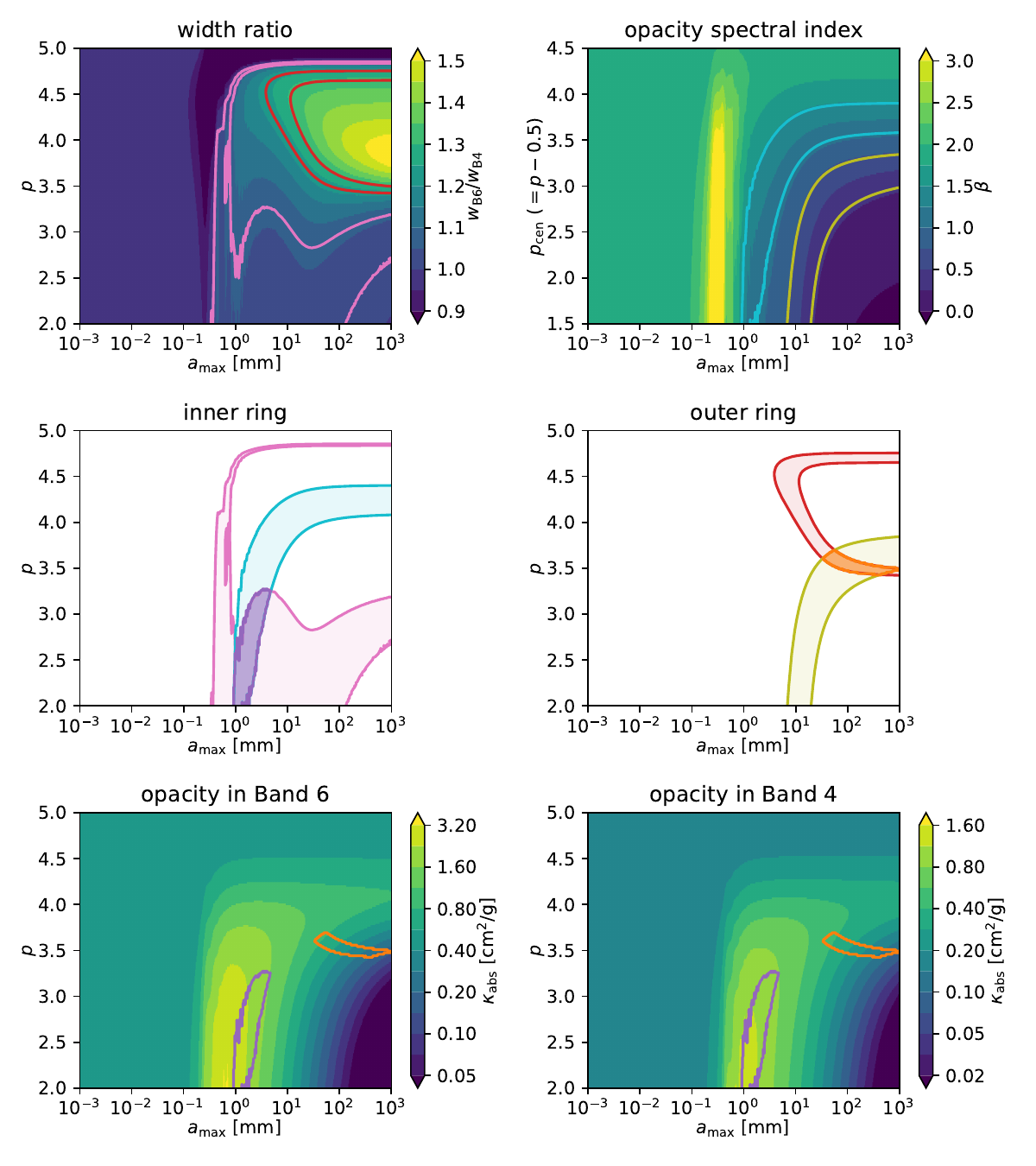}
        \caption{
            Top left: model ring width ratio assuming size dependent dust trapping for given dust size $a_{\mathrm{max}}$ and exponent of dust size distribution $p$. 
            The orange contour shows observed constraint in the inner ring, and the red one shows that in the outer ring.
            Top right: model opacity spectral index $\beta$.
            Middle left and right: constrained parameter space of $a_{\mathrm{max}}$ and $p$ in the inner ring (left) and in the outer ring (right) from the ring width ratio and the spectral index.
            Bottom left and right: dust opacity in Band 6 (left) and in Band 4 (right).
       \label{fig:width_ratio}}
    \end{center}
\end{figure*}

Next, we investigate the dependence of the ring width ratio between Band 6 (1.25 mm) and Band 4 (2.14 mm) on the dust size distribution.
The top left panel in Figure \ref{fig:width_ratio} shows the ring width ratio between Band 6 and Band 4 for different maximum dust sizes $a_{\mathrm{max}}$ and exponents $p$.
To calculate this ratio, we fit the model optical depth profile with a Gaussian function for regions where $\tau_{\mathrm{normalize}} > 0.5$.
We find that the dust ring width ratio is larger than 1.2 only if the exponent $3.4 < p < 4.8$ and the maximum dust size $a_{\mathrm{max}} > 3\ \mathrm{mm}$.
The dark blue region shows that the dust ring width is narrower in Band 6 than in Band 4 because the largest grains in these dust size distributions correspond to the opacity peak in Figure \ref{fig:opacity} only in Band 6, resulting in the optical depth distribution being weighted towards the larger side of the grain distribution and appearing narrower in Band 6.
Thus, the dust width ratio depends on the dust size distribution.
Conversely, we can constrain the dust size distribution from the observed dust ring width ratio.

\subsection{Opacity spectral index} \label{sec:model_sed}

We investigate the dependence of the opacity spectral index on the dust size distribution.
We calculate the optical depth at the ring center using equation (\ref{eq:opacity_distribution}) with $\Delta r = 0$ and calculate the opacity spectral index using equation (\ref{eq:def_beta}).
We note that the exponent of the dust size distribution at the ring center $p_{\mathrm{cen}}$ is related to the exponent of the dust size distribution in the entire ring $p$ as $p_{\mathrm{cen}} = p - 0.5$ due to the size-dependent dust trapping.

The top right panel of Figure \ref{fig:width_ratio} shows the opacity spectral index for each maximum dust size $a_{\mathrm{max}}$ and exponent $p$.
If the maximum dust size is small, such as $a_{\mathrm{max}} < 100\ \mathrm{\mu m}$, or small grains contribute to the dust opacity with $p > 4.0$, the opacity spectral index $\beta$ has a constant value of 1.7. 
If the maximum dust size is around a few hundred micrometers, the opacity spectral index reaches maximum values due to the opacity peak. 
The larger maximum dust size $a_{\mathrm{max}}$ and smaller exponent $p$ correspond to the smaller $\beta$.

\section{Constraint on the dust size distribution of the HD 163296 disk} \label{sec:dust_size}

We constrain the dust size distribution of the HD 163296 disk by comparing the observations in Section \ref{sec:obs} and the physical models in Section \ref{sec:modeling}.
We constrain the maximum dust size $a_{\mathrm{max}}$ and the exponent $p$ from the ring width ratio in Section \ref{sec:width} and from the opacity spectral index in Section \ref{sec:sed}, and then derive the final constraints on dust size distribution from the intersection of these constraints in Section \ref{sec:param_space}.

\subsection{Constraints on the dust size distribution from the dust ring width ratio} \label{sec:width}

First, we constrain the dust size distribution from the perspective of the ring width ratio between Band 6 and Band 4.
In Section \ref{sec:obs}, we estimated that the ring width ratio for the inner ring is $w_{\mathrm{B6}}/w_{\mathrm{B4}} = 1.04^{+0.03}_{-0.03}$, while for the outer ring, it is $w_{\mathrm{B6}}/w_{\mathrm{B4}} = 1.27^{+0.04}_{-0.04}$.
From the consistency between the observations and the physical models, the parameter space of the dust size distribution in the inner ring is constrained to the pink contoured region in the top left panel of Figure \ref{fig:width_ratio} and that in the outer ring is constrained to the red contoured region.

\subsection{Constraints on the dust size distribution from the spectral index} \label{sec:sed}

First, we qualitatively compare the radial profile of the observed slope of the spectral energy distribution $\alpha$ with the model.
The top panel of Figure \ref{fig:SED} shows the radial intensity profile in Band 4 and Band 6, and the bottom panel shows the observed spectral slope $\alpha$ between the two Bands.
We smooth the Band 6 image to have the same resolution as the Band 4 image.
We can see that the SED has a local minimum at the ring center as shown in the bottom panel of Figure \ref{fig:model_radial}, which suggests that larger dust accumulates at the ring center as pointed out by \citet{Sierra2021}.
However, the spectral slope $\alpha$ depends not only on the opacity spectral index $\beta$ (dust size) but also on the optical depth.
Therefore, we need to consider the optical depth effects to derive the dust size properties.

We compare the opacity spectral index $\beta$ from the fitting of the observations at the ring center with the model to quantitatively constrain the dust size distribution.
From the parameters listed in Table \ref{tab:fitted}, we derive $\beta_{\mathrm{cen}} = 1.27^{+0.05}_{-0.05}$ for the inner ring and $\beta_{\mathrm{cen}} = 0.51^{+0.04}_{-0.05}$ for the outer ring. 
Accounting for flux calibration error, we derive $\beta_{\mathrm{cen}} = 1.27^{+0.21}_{-0.21}$ for the inner ring and $\beta_{\mathrm{cen}} = 0.51^{+0.21}_{-0.21}$ for the outer ring.
By comparing the opacity spectral index from the observations with the model presented in the top right panel of Figure \ref{fig:width_ratio}, the parameter space of the dust size distribution is constrained to the cyan contoured region for the inner ring, and to the yellow contoured region for the outer ring.

\subsection{Dust size distribution satisfies these 2 conditions} \label{sec:param_space}

Combining the two constraints, we obtain the final results of the parameter space that satisfies both the wavelength dependence of the spatial distribution and the opacity spectral index.
We constrain the parameter space shown in the middle panels of Figure \ref{fig:width_ratio} where
$9 \times 10^{-1}\ \mathrm{mm } < a_{\mathrm{max}} <  5\ \mathrm{mm}$ and $p < 3.3$ in the inner ring and $3 \times 10^1 \ \mathrm{mm} < a_{\mathrm{max}} < 1 \times 10^3$ and $3.4 < p < 3.7$ in the outer ring.

We note that the assumed dust spatial distribution in equation (\ref{eq:radial_approx-2}) is not valid for dust grains with $\mathrm{St} > 1$, which corresponds to $a > 10^2 \ \mathrm{mm}$ given the gas surface density assumed in Section \ref{sec:turbulence}.
Consequently, the upper limit of the dust size in the outer ring cannot be reliably determined.
Therefore, we only consider the lower limit of $3 \times 10^1 \ \mathrm{mm} < a_{\mathrm{max}}$ in the outer ring for further discussions.

\subsection{Comparison with previous studies} \label{sec:comparison}

There are several previous studies that estimate the dust size distribution of the HD 163296 disk from multiwavelength observations \citep{Guidi2016,Guidi2022,Sierra2021}.
\citet{Sierra2021} estimated the dust surface density and the dust size from ALMA Band 6 (1.3 mm) and Band 3 (3.0 mm) observations with intermediate resolution ($\sim 0^{\prime\prime}.2$) with the temperature model by \citet{Zhang2021_MAPS5}.
They derived the maximum dust size $a_{\mathrm{max}}$ to be $a_{\mathrm{max}} \sim 2\ \mathrm{mm}$ at the inner ring and $a_{\mathrm{max}} \sim 3\ \mathrm{mm}$ at the outer ring, assuming the dust temperature and fixing the power law exponent $p$ to be 2.5.
They also found that the dust size is larger at the ring center only in the outer ring.
The estimated dust size in the inner ring is consistent with our results, but that in the outer ring is smaller than our results.
The difference may come from the fact that their observations do not resolve the ring and that they fixed the power law exponent $p$ to be 2.5.

\citet{Guidi2022} estimated the dust temperature, surface density, exponent of the size distribution, and dust size from ALMA Band 7 (0.9 mm), 6 (1.3 mm), 4 (2.1 mm), 3 (3.0 mm), and VLA Ka Band (9.1 mm) observations without assuming the dust temperature.
They obtained the high resolution ($\sim 0.^{\prime\prime}09$) radial profiles by visibility fitting using \texttt{frank} \citep{Jennings2020} assuming axisymmetry while ignoring the non-axisymmetric structures.
They found that the dust size is 100 $\mathrm{\mu m}$ across the disk, and there is no difference in the dust size between the ring and gaps.

One of the reasons for the difference between our results and those of \citet{Guidi2022} is the difference in the dust model.
We use the DSHARP dust model, which assumes organics as a carbonaceous material \citep{Henning1996}, whereas \citet{Guidi2022} adopt the DIANA dust model \citep{Min2016}, which assumes amorphous carbon \citep{Zubko1996}.
We present results using different dust models in Appendix \ref{sec:porous}. 
With a dust model including amorphous carbon, we derive smaller dust sizes, which are closer to those reported by \citet{Guidi2022}.
However, while the discrepancy in estimated dust sizes decreases when using a similar dust model, it cannot be solely attributed to the dust model.

The inconsistencies may also arise from differing assumptions regarding the dust spatial and size distribution. 
\citet{Guidi2022} assumed a power-law dust size distribution, exploring four different slopes at each radius. 
In contrast, our model presumes a power-law dust size distribution over the entire ring, but the local dust size distribution does not follow power law due to the size-dependent dust trapping.

Our estimation has the following advantages and disadvantages compared to that of \citet{Guidi2022}.
First, we solely use the Band 6 and Band 4 images, which have the highest spatial resolutions ($\sim 0.^{\prime\prime}06$), a subset of the images also used by \citet{Guidi2022}.
These spatially resolving observations make it possible to find the variation of the dust size across the ring by using the observations that resolve the ring ($\sim 0.^{\prime\prime}1$ in FWHM).
However, we use fewer wavelengths than \citet{Guidi2022} and need to assume the temperature model.
Second, we assume the Gaussian optical depth models to fit the observations, while \citet{Guidi2022} used \texttt{frank} to make the non-parametric radial profile.
\citet{Guidi2022} estimated the physical properties that reproduce the smoothed intensity profiles, and thus, their results are affected by the angular resolutions.
On the other hand, we estimate the physical properties from the optical depth profile before smoothing using forward modeling.
While this method allows us to derive physical properties without a smoothing effect, it has model dependence.

\section{Physical properties of the HD 163296 disk} \label{sec:dis}

We discuss the physical properties of the disk from the dust spatial distribution and the dust size distribution in Section \ref{sec:obs} and \ref{sec:dust_size}.
In Section \ref{sec:interp}, we discuss the interpretations of the estimated maximum dust size and the exponent.
In Section \ref{sec:a_St}, we constrain $\alpha / \mathrm{St}$ from the dust spatial distribution estimated in \ref{sec:fitting_res} by assuming the gas spatial distribution from the literature. 
In Section \ref{sec:turbulence}, we constrain the gas turbulence strength $\alpha$ from $\alpha/\mathrm{St}$ and the dust size estimated in Section \ref{sec:dust_size}.
Thereafter, we individually discuss the properties of the disk or the dust grains based on the derived physical quantities.

\subsection{Interpretation of the dust size distribution} \label{sec:interp}

The maximum dust size in the outer ring ($3 \times 10^1 \ \mathrm{mm} < a_{\mathrm{max}}$) is larger than that in the inner ring ($9 \times 10^{-1}\ \mathrm{mm } < a_{\mathrm{max}} <  5\ \mathrm{mm}$).
This result is opposite to the theoretical expectation that the dust grains are larger at the inner side because of the shorter growth timescale \citep[][]{Brauer2008}.
It suggests that the dust growth is locally suppressed in the inner ring and/or locally enhanced in the outer ring.
The suppression of the dust growth can be caused by enhanced fragmentation due to sintering around the snowlines \citep{Okuzumi2016,Sirono2017}, and the inner ring radius is consistent with the CO snow line based on the observations \citep{Zhang2021_MAPS5}.
These processes may affect the efficiency or the location of the planetesimal formation.

The exponent $p \sim 3.5$ in the outer ring is consistent with the collisional cascade of larger bodies \citep[][]{Dohnanyi1969,Tanaka1996}.
It may imply that the planetesimals are already formed around the outer ring.
We further discuss the equilibrium between dust growth and fragmentation in Section \ref{sec:frag_pow}.

\subsection{\texorpdfstring{$\alpha / \mathrm{St}$}{a/St}} \label{sec:a_St}

We constrain the ratio of $\alpha_{\{r,z\}}$ to $\mathrm{St}$ from the dust spatial distributions.
In the equilibrium between dust diffusion and drift/settling in a Gaussian gas ring with the gas ring width $w_{\mathrm{gas}}$ and the gas scale height $h_{\mathrm{gas}}$, the dust ring width $w_{\mathrm{dust}}$ and the dust scale height $h_{\mathrm{dust}}$ are given by
\begin{equation} \label{eq:radial}
    w_{\mathrm{dust}} = \left( 1 + \frac{\mathrm{St}}{\alpha_r} \right)^{-1/2} w_{\mathrm{gas}},
\end{equation}
\begin{equation} \label{eq:vertical}
    h_{\mathrm{dust}} = \left( 1 + \frac{\mathrm{St}}{\alpha_z} \right)^{-1/2} h_{\mathrm{gas}}.
\end{equation}
We constrain $\alpha_{\{r,z\}}/\mathrm{St}$ using equations (\ref{eq:radial}) and (\ref{eq:vertical}) from the dust and gas ring width.
Since the dust grains of different sizes have different spatial distribution, $\alpha/\mathrm{St}$ depends on the dust size in focus.
Here, we derive $\alpha/\mathrm{St}$ from the optical depth distribution fitted in Section \ref{sec:fitting_res}. 
In other words, we derive $\alpha/\mathrm{St}$ for the dust grains that dominate the optical depth.

We use the gas ring width derived by \citet{Rosotti2020}, who estimated the gas ring width based on the deviation of the CO line emission observed by ALMA from the Keplerian rotation.
The resulting gas ring widths are $w_{\mathrm{gas}} = 14.4 \pm 1.0 \ \mathrm{au}$ at the inner ring and $w_{\mathrm{gas}} = 23.2 \pm 1.3\ \mathrm{au}$ at the outer ring.

We assume the gas scale height assuming hydrostatic equilibrium in the vertical direction.
We assume the central stellar mass $M_{\mathrm{star}} = 2.02 M_{\mathrm{sum}}$ from \citet{Teague2019_Nat} and the temperature profile as equation (\ref{eq:temperature}).
The derived gas scale height is $h_{\mathrm{gas}} = 4.07\ \mathrm{au}$ at the inner ring and $h_{\mathrm{gas}} = 6.65\ \mathrm{au}$ at the outer ring.

We summarize the derived $\alpha_{\{r,z\}} / \mathrm{St}$ in Table \ref{tab:a_St}.
For the radial direction, we derive $\alpha_{r} / \mathrm{St} = 1.3^{+0.2}_{-0.2}\times10^{-1}$ in Band 6 and $\alpha_{r} / \mathrm{St} = 1.2^{+0.2}_{-0.2}\times10^{-1}$ in Band 4 in the inner ring, 
and $\alpha_{r} / \mathrm{St} = 4.2^{+0.4}_{-0.4}\times10^{-2}$ in Band 6 and  $\alpha_{r} / \mathrm{St} = 2.6^{+0.3}_{-0.2}\times10^{-2}$ in Band 4 in the outer ring.
For the vertical direction, we derive $\alpha_{z} / \mathrm{St} = 5.2^{+7.5}_{-2.1}$ in Band 4 and in the inner ring, 
and $\alpha_{z} / \mathrm{St} < 7.3\times10^{-3}$ in Band 6 and $\alpha_{z} / \mathrm{St} < 2.1\times10^{-2}$ in Band 4 in the outer ring.
We confirm that $\alpha/\mathrm{St}$ is larger in the inner ring than in the outer ring in both radial and vertical directions, as shown in \citet[][]{Rosotti2020, Doi2021}.
We further discuss the anisotropy of the turbulence in Section \ref{sec:turb_mech}. 
We cannot constrain $\alpha_z/\mathrm{St}$ in the inner ring in Band 6 because $h_{\mathrm{dust}} > h_{\mathrm{gas}}$, which is not physical.
While we assume axisymmetry in this study, we note the disk may have non-axisymmetric structures, resulting in an overestimation of the dust scale height.
We may also underestimate the gas scale height due to the assumption of the temperature.

\subsection{gas turbulence} \label{sec:turbulence}

We estimate the turbulence parameter $\alpha$ from $\alpha/\mathrm{St}$ constrained in Section \ref{sec:a_St} with an assumption of Stokes number $\mathrm{St} (= \pi \rho_{\mathrm{mat}} a / 2 \Sigma_{\mathrm{gas}})$.
We adopt the dust material density $\rho_{\mathrm{mat}} = 1.675\ \mathrm{g/cm^3}$ from the DSHARP dust model \citep[][]{Birnstiel2018} assuming compact grains.
We adopt the gas surface density at the inner ring as $\Sigma_{\mathrm{gas}} = 21.1 \ \mathrm{g/cm^2}$ and at the outer ring as $\Sigma_{\mathrm{gas}} = 10 \ \mathrm{g/cm^2}$ based on the modeling of CO isotopologue observations using ALMA \citep{Rab2020}.

To estimate the Stokes number, we need to determine a single, representative dust size, even though dust grains have a size distribution. 
We define the representative dust size for each ring using the following approach. 
First, we model the optical depth profile assuming a power-law dust size distribution as in Section \ref{sec:model_profile}. 
Then, we fit the model profile with a Gaussian function and define the representative dust size as the size of single-sized dust grains that have the same ring width as the fitted Gaussian.
Here, we assume an exponent of $p=3.5$, as estimated in Section \ref{sec:dust_size}.

\begin{figure}
    \begin{center}
        \includegraphics[width=8 cm]{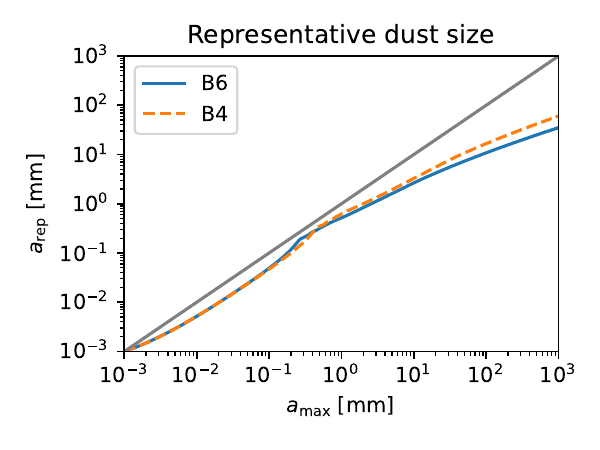}
        \caption{Representative dust size if $p = 3.5$ in Band 6 (blue) and Band 4 (orange).
        The gray line shows $a_{\mathrm{rep}} = a_{\mathrm{max}}$.
        \label{fig:traced_size}}
    \end{center}
\end{figure}

Figure \ref{fig:traced_size} shows representative dust size for a given maximum dust size.
The representative dust size is always smaller than the maximum dust size because the dust opacity distribution is a superposition of the dust opacity distributions from the minimum to maximum dust sizes. 
For example, when we assume $a_{\mathrm{max}} = 2\ \mathrm{mm}$ and $p=3.5$ within the constraint for the inner ring, 
the representative dust size is $a_{\mathrm{rep}} = 0.8\ \mathrm{mm}$ in Band 6 and $a_{\mathrm{rep}} = 1\ \mathrm{mm}$ in Band 4. 
When $a_{\mathrm{max}} > 3 \times 10^1\ \mathrm{mm}$ and $p=3.5$ as constrained in the outer ring, the representative dust size is $a_{\mathrm{rep}} > 5\ \mathrm{mm}$ in Band 6 and $a_{\mathrm{rep}} > 8\ \mathrm{mm}$ in Band 4.
These representative dust size are adopted in the following discussion.
The following discussion is a rough estimation, and $\mathrm{St}$ is proportional to the assumed $a_{\mathrm{rep}}$ and inversely proportional to the assumed gas surface density.

We calculate the Stokes number and constrain the turbulence parameter $\alpha_{{r,z}}$.
Our results are summarized in Table \ref{tab:a_St}.
From the observations in both Bands, we derive the radial turbulence strength
$\alpha_r \sim 1 \times 10^{-3}$ in the inner ring, and $\alpha_r > 5 \times 10^{-3}$ in the outer ring.
We derive the vertical turbulence strength  $\alpha_z \sim 6 \times 10^{-2}$ from the Band 4 observations in the inner ring.
In the outer ring, we cannot put constraints on $\alpha_z$ from the upper limit of $\alpha / \mathrm{St}$ and the lower limit of St.
The vertical turbulence strength in the inner ring is much stronger than theoretical expectation \citep[e.g.,][]{Fukuhara2022,Xu2022}.

\begin{longrotatetable}
    \tabletypesize{\scriptsize}
\begin{deluxetable*}{ccccccccccccccc}
        \tablecaption{$\alpha/\mathrm{St}$\label{tab:a_St}}
        \tablewidth{0pt}
        \tablehead{\colhead{ring} & \colhead{$\Sigma_{\mathrm{gas}}$} & \colhead{wavelength} & \colhead{$a_{\mathrm{max}}$} & \colhead{$a_{\mathrm{rep}}$} & \colhead{St} & \colhead{$w_{\mathrm{gas}}$} & \colhead{$w_{\mathrm{dust}}$}  & \colhead{$h_{\mathrm{gas}}$} & \colhead{$h_{\mathrm{dust}}$} & \colhead{$\alpha_r / \mathrm{St}$} & \colhead{$\alpha_z / \mathrm{St}$} & \colhead{$\alpha_z/\alpha_r$} & \colhead{$\alpha_r$} & \colhead{$\alpha_z$} \\
        \colhead{} & \colhead{[$\mathrm{g/cm^2}$]} & \colhead{[mm]} & \colhead{[mm]} & \colhead{[mm]} &   & \colhead{[au]} & \colhead{[au]} & \colhead{[au]} & \colhead{[au]} \\
        \colhead{Ref.} & \colhead{(1)} & \colhead{} & \colhead{} & \colhead{} & \colhead{} & \colhead{(2)} & \colhead{} & \colhead{(3)}}
        \startdata
        \multirow{2}{*}{inner} & \multirow{2}{*}{21.0} 
        & 1.25
        & \multirow{2}{*}{$0.9 - 5$} 
        & $0.8$ & $1\times10^{-2}$ 
        & \multirow{2}{*}{$14.4^{+1.0}_{-1.0}$} 
        & $4.82^{+0.08}_{-0.08}$
        & \multirow{2}{*}{$4.07$}
        & $4.28^{+0.13}_{-0.13}$ & $1.3^{+0.2}_{-0.2}\times10^{-1}$ & $-$ & $-$ & $1\times10^{-3}$ & $-$
        \\
        &
        & 2.14
        &
        & $1$ & $1\times10^{-2}$ 
        &
        & $4.65^{+0.12}_{-0.11}$
        &
        & $3.75^{+0.19}_{-0.20}$ & $1.2^{+0.2}_{-0.2}\times10^{-1}$ & $5.2^{+7.5}_{-2.1}$ & $4.4^{+6.4}_{-2.0} \times 10^1$ & $1\times10^{-3}$ & $6\times10^{-2}$
        \\
        \multirow{2}{*}{outer} & \multirow{2}{*}{10.0}
        & 1.25
        & \multirow{2}{*}{$> 3 \times 10^1$} 
        & $> 5$ & $>1\times10^{-1}$ 
        & \multirow{2}{*}{$23.2^{+1.3}_{-1.3}$} 
        & $4.66^{+0.09}_{-0.09}$
        & \multirow{2}{*}{$6.65$}
        & $<0.57$ & $4.2^{+0.4}_{-0.4}\times10^{-2}$ & $<7.3\times10^{-3}$ & $<1.9 \times 10^{-1}$ & $>5\times10^{-3}$ & $-$
        \\
        &
        & 2.14
        &
        & $> 8$ & $>2\times10^{-1}$
        &
        & $3.68^{+0.09}_{-0.10}$
        &
        & $<0.97$ & $2.6^{+0.3}_{-0.2}\times10^{-2}$ & $<2.1\times10^{-2}$ & $< 9.2 \times 10^{-1}$ & $>5\times10^{-3}$ & $-$
        \enddata
        \tablecomments{Estimated dust size, $\alpha/\mathrm{St}$, and $\alpha$. \\
        References are (1) \citet{Rab2020}; (2) \citet{Rosotti2020}; (3) calculated using the central stellar mass from \citet{Flaherty2015} and the temperature profile from \citet{Isella2016} assuming hydrostatic equilibrium.}
\end{deluxetable*}
\end{longrotatetable}

\subsection{anisotoropy of the turbulence} \label{sec:turb_mech}

We discuss the anisotropy of gas turbulence and its mechanisms.
Since gas turbulence can be anisotropic depending on the mechanisms, observational constraints on turbulence anisotropy provide clues to constrain the mechanism of turbulence.
We compare $\alpha_r/\mathrm{St}$ and $\alpha_z/\mathrm{St}$ and constrain the turbulence anisotropy $\alpha_z / \alpha_r$ to be $\alpha_z/\alpha_r = 4.4^{+6.4}_{-2.0} \times 10^{1}$ from the Band 4 observations in the inner ring, and $\alpha_z/\alpha_r < 1.9 \times 10^{-1}$ from the Band 6 observations and $\alpha_z/\alpha_r < 9.5 \times 10^{-1}$ from the Band 4 observations in the outer ring.
In the inner  ring, the turbulence is stronger in the vertical direction, while in the outer ring, it is similar or slightly stronger in the radial direction.

Our findings of different turbulence strengths and anisotropies at different regions suggest that the mechanism of gas turbulence differs between the two rings.
A possible mechanism that causes stronger diffusion in the vertical direction, as in the inner ring, is vertical shear instability (VSI).
\citet{Dullemond2022} shows that VSI makes a thin and corrugated dust layer.
However, the dust layer appears vertically thick layer if the radial wavelength of VSI is smaller than the ring width \citep{Pfeil2021}.
Also, the dust layer can be thick if we assume the dust size distribution since the dust elevation height depends on the dust size, while they assume single size dust grains.
Another possible mechanism to puff up the dust vertically is the meridional flows generated by the planets \citep[][]{Teague2019_Nat,Bi2021}.

In the outer ring, the turbulence is almost isotropic with small excess for radial direction.
Magneto-rotational instability (MRI) provokes isotropic or slightly stronger turbulence in the radial direction \citep{Flock2011, Suzuki2014}, as in the outer ring. 
Convective overstability (COV) also provokes radially dominant turbulence \citep[e.g.,][]{Lyra2014,Raettig2021}.
Another explanation for radially prominent diffusion is that a planet makes additional diffusion as a perturber and provokes radially diffused dust distribution \citep[][]{Bi2022arxiv}.

\subsection{dust size distribution of fragments at a destructive collision} \label{sec:frag_pow}

We constrain the dust fragmentation properties based on the derived exponent of the dust size distribution, assuming equilibrium between collisional growth and fragmentation.
We assume that dust grains are trapped in the dust ring and do not take into account mass loss due to radial drift.
Additionally, we assume that the spatial distribution of dust grains is determined by the equilibrium between accumulation and diffusion and do not consider collisional trapping of small grains \citep[][]{Krijt2016}.

\citet{Birnstiel2011} analytically derived a relationship between the exponent of the dust size distribution and the fragments as
\begin{equation}
p = \frac{p_{\gamma} + p_{\epsilon} + 3}{2},
\end{equation}
where $p$ is the exponent of the dust size distribution in the disk, $p_{\epsilon}$ is the exponent of the fragments, and $p_{\gamma}$ is the exponent of the collision kernel.
Assuming the dust relative velocity is determined by gas turbulence as $\Delta v \propto a^{1/2}$, and the dust radial and vertical distribution follows $w_{\mathrm{d}} \propto h_{\mathrm{d}} \propto a^{-1/2}$ due to the size-dependent dust trapping, the collisional kernel is $p_{\gamma} = 3/2$. 
Since $p \sim 3.5$ in the outer ring, we can constrain $p_{\epsilon} \sim 2.5$.
This constraint is steeper than the experimental results of $p_{\epsilon}$ ranging from 1.2 to 2.1 \citep[][]{Blum1993,Guttler2010} and the numerical simulation of collisions of dust aggregates \citep{Wada2009, Hasegawa2022arXiv}.
However, it is flatter than the observed constraint of $p = 3.5$ for the interstellar medium (ISM) \citep{Mathis1977}, which is often used as the dust size distribution of fragments as $p=p_{\epsilon}$ in simulations of global disk evolution \citep{Brauer2008, Stammler2022}.

\subsection{dust fragmentation velocity}

We constrain the dust fragmentation velocity $v_{\mathrm{frag}}$ from the constrained maximum dust size and the gas turbulence assuming the fragmentation limit.
The fragmentation velocity driven by the gas turbulence is given as
\begin{equation}
    \Delta v_t \sim \sqrt{\alpha \mathrm{St}} c_s.
\end{equation}
If dust growth is limited by collisional fragmentation, the relative velocity $\Delta v$ is equal to the fragmentation velocity $v_{\mathrm{frag}}$.
We adopt the maximum dust size as $0.9 \ \mathrm{mm} < a_{\mathrm{max}} < 5 \ \mathrm{mm}$ in the inner ring and $3 \times 10^1 \ \mathrm{mm} < a_{\mathrm{max}}$ in the outer ring, as derived in Section \ref{sec:dust_size}.
We assume that the largest component of the gas turbulence determines the dust relative velocity and use $\alpha_z > 6 \times 10^{-2}$ in the inner ring and $\alpha_r \sim 5 \times 10^{-3}$ in the outer ring. 
We assume the gas surface density in Section \ref{sec:turbulence} and the temperature in Section \ref{sec:a_St}.
Then, we constrain that $ 8 \ \mathrm{m/s} < v_{\mathrm{frag}} < 2 \times 10^1\ \mathrm{m/s}$ in the inner ring and $ 2 \times 10^1 \ \mathrm{m/s} < v_{\mathrm{frag}}$ in the outer ring. 
Laboratory experiments \citep{Gundlach2018, Musiolik2019,Steinpilz2019} and numerical calculations \citep[][]{Wada2009} of dust collision show that smaller monomer size and larger surface energy result in higher fragmentation velocity.
The high fragmentation velocity shown in this study suggests small monomer size and/or large surface energy.

\subsection{dust mass}

We estimate the dust mass of the two rings from the observed intensity.
We integrate the optical depth distribution in Equation (\ref{eq:distribution}) and derive the dust mass in a ring as
\begin{equation}
M = 2 \pi^{3/2} \sin{(i)} \tau_0 r_0 w_{\mathrm{dust}} /\kappa.
\end{equation}
The dust opacity as a function of the maximum dust size and the exponent is shown in the bottom panels of Figure \ref{fig:width_ratio}.
Within the constrained parameter range of the dust size distribution, the opacity $\kappa_{\mathrm{abs}} < 2.9 \ \mathrm{g/cm^{-2}}$ in the inner ring and $\kappa_{\mathrm{abs}} < 0.88\ \mathrm{g/cm^{-2}}$ in the outer ring in Band 6. 
Therefore, we put lower limits for the dust mass in each ring using the ring parameters fitted in Section \ref{sec:obs} and derive $m > 2.0 \times 10^{{29}}\ \mathrm{g} = 34 M_{\mathrm{earth}}$ in the inner ring and $m > 5.8 \times 10^{{29}}\ \mathrm{g} = 97 M_{\mathrm{earth}}$ in the outer ring.
The derived lower limit of the dust masses are enough to form solid planets in both rings.
Also, the lower limit of the dust mass is consistent with those of the inner ring ($53^{+11}_{-9} M_{\mathrm{earth}}$) and the outer ring ($96^{+13}_{-16} M_{\mathrm{earth}}$) estimated by \citet{Guidi2022} and the total disk mass ($2.8^{+4.5}_{-0.6} \times 10^2 M_{\mathrm{earth}} (= 0.83^{+1.35}_{-0.18} \times 10^{-3} M_{\mathrm{sun}})$) estimated by \citet{Sierra2021}.

\section{Conclusion} \label{sec:concl}

We analyzed ALMA high-resolution dust continuum images of the HD 163296 disk in Band 6 (1.25 mm) and Band 4 (2.14 mm).
We fit the two rings at 67 au and 100 au in the two bands individually, assuming Gaussian optical depth distributions. 
We discuss the dust size distribution from two aspects: the difference in the dust spatial distribution between the two wavelengths and the opacity spectral index between the two wavelengths.
We also discussed the physical properties of the disk from the dust distributions.
Our major conclusions are as follows.

\begin{enumerate}
    \item We found that the dust ring appears narrower at the longer wavelength in the outer ring (100 au).
    In contrast, the dust ring appears to have similar widths in the inner ring (67 au). 
    The ring width ratio of the optical depth distribution in the inner ring is $w_{\mathrm{B6}}/w_{\mathrm{B4}} = 1.04^{+0.03}_{-0.03}$, while in the outer ring it is $w_{\mathrm{B6}}/w_{\mathrm{B4}} = 1.27^{+0.04}_{-0.04}$. 
    The wavelength dependence in the outer ring is consistent with the dust-trapping model.
    \item We constrained the dust size distributions characterized by the maximum dust size $a_{\mathrm{max}}$ and the power law exponent $p$ from the wavelength dependence of the ring width and the opacity spectral index between the two Bands. 
    We find that $0.9 \ \mathrm{mm} < a_{\mathrm{max}} < 5 \ \mathrm{mm}$ and $p < 3.3$ in the inner ring, and $a_{\mathrm{max}} > 3 \times 10^1 \ \mathrm{mm}$ and $3.4 < p < 3.7$ in the outer ring. 
    These different maximum dust sizes suggest that the degree of dust growth varies locally, and planetesimals are formed locally. The exponent $p \sim 3.5$ may imply the collisional cascade of already formed planetesimals in the outer ring.
    \item We constrained the $\alpha/\mathrm{St}$ from the dust spatial distribution.
    We compared $\alpha_r/\mathrm{St}$ and $\alpha_z/\mathrm{St}$ and found that in the inner ring, the vertical diffusion dominates over the radial diffusion, while in the outer ring, the diffusion is almost isotropic with slight excess for the radial direction.
    It suggests that different turbulence mechanisms may be operating between the two rings.
    \item We constrained $\alpha$ from $\alpha/\mathrm{St}$ and the dust size.
    We derived that in the radial direction, $\alpha_r \sim 1 \times 10^{-3}$ in the inner ring, and $\alpha_r > 5 \times 10^{-3}$ in the outer ring at both wavelengths.
    We derived that in the vertical direction, $\alpha_z > 6 \times 10^{-3}$ in Band 4 in the inner ring. 
    \item We constrained the dust fragmentation velocity from the constrained maximum dust size and the gas turbulence assuming that the dust size is limited by collisional fragmentation by the gas turbulence.
    We found that $ 8 \ \mathrm{m/s} < v_{\mathrm{frag}} < 2 \times 10^1\ \mathrm{m/s}$ in the inner ring and $ 2 \times 10^1 \ \mathrm{m/s} < v_{\mathrm{frag}}$ in the outer ring. 
    \item We discussed the exponent of the dust size distribution of fragments.
    We assumed the equilibrium between the collisional coagulation and fragmentation and constrain the exponent of the fragments $p_f$ produced by a destructive collision.
    We constrained $p_f \sim 2.5$ in the outer ring.

\end{enumerate}

\section*{Acknowledgements} 

The authors thank the anonymous referee for their helpful comments and suggestions that significantly improved the manuscript.
The authors thanks for T. Birnstiel and S. Stammler for their helpful comments for the code improvement.
This work was supported by JSPS KAKENHI Grant Numbers JP18K13590, JP22K03680, JP22J20739, and JP22KJ1435.
This paper makes use of the following ALMA data: ADS/JAO.ALMA\#2013.1.00366.S, ADS/JAO.ALMA\#2013.1.00601.S, ADS/JAO.ALMA\#2016.1.00484.L, and ADS/JAO.ALMA\#2017.1.01682.S. 
ALMA is a partnership of ESO (representing its member states), NSF (USA) and NINS (Japan), together with NRC (Canada), MOST and ASIAA (Taiwan), and KASI (Republic of Korea), in cooperation with the Republic of Chile. The Joint ALMA Observatory is operated by ESO, AUI/NRAO and NAOJ.
The authors thank the DSHARP team and \citet{Guidi2022} for making their data publicly available.

Facility: ALMA

Software:
\texttt{astropy}\citep{astropy2022}
\texttt{emcee}\citep{Foreman-Mackey2013},
\texttt{scipy}\citep{scipy2020}

\appendix

\begin{deluxetable*}{cccccccccc}
        \tablecaption{fitted parameters\label{tab:fitted_app}}
        \tablewidth{0pt}
        \tablehead{\colhead{model} & \colhead{ring} & \colhead{wavelength} & \colhead{$r_0$} & \colhead{$\tau_0$} & \colhead{$w_0$} & \colhead{$h_0$} & \colhead{$x_{\mathrm{cen}}$} & \colhead{$y_{\mathrm{cen}}$} & \colhead{$w_{\mathrm{B6}}/w_{\mathrm{B4}}$}\\
        \colhead{} & \colhead{} & \colhead{[mm]} & \colhead{[au]} & \colhead{} & \colhead{[au]} & \colhead{[au]} & \colhead{[au]} & \colhead{[au]}}
        \startdata
    \multirow{4}{*}{Fiducial}
        & \multirow{2}{*}{inner}  
        & 1.25 & $67.82^{+0.05}_{-0.05}$ & $0.718^{+0.011}_{-0.011}$ & $4.82^{+0.08}_{-0.08}$ & $4.28^{+0.13}_{-0.13}$ & $0.80^{+0.06}_{-0.05}$ & $1.19^{+0.06}_{-0.06}$ & \multirow{2}{*}{$1.04^{+0.03}_{-0.03}$} \\
        & & 2.14 & $67.79^{+0.07}_{-0.07}$ & $0.363^{+0.008}_{-0.007}$ & $4.65^{+0.12}_{-0.11}$ & $3.75^{+0.19}_{-0.20}$ & $1.08^{+0.07}_{-0.07}$ & $1.26^{+0.07}_{-0.08}$ \\
        & \multirow{2}{*}{outer}  
        & 1.25 & $100.41^{+0.06}_{-0.06}$ & $0.423^{+0.006}_{-0.005}$ & $4.66^{+0.09}_{-0.09}$ & $<0.57(0.28^{+0.29}_{-0.19})$ & $0.48^{+0.08}_{-0.08}$ & $1.30^{+0.08}_{-0.08}$ & \multirow{2}{*}{$1.27^{+0.04}_{-0.04}$}  \\
        & & 2.14 & $100.45^{+0.06}_{-0.06}$ & $0.322^{+0.007}_{-0.006}$ & $3.68^{+0.09}_{-0.10}$ & $<0.97(0.51^{+0.47}_{-0.35})$ & $0.85^{+0.09}_{-0.08}$ & $0.85^{+0.10}_{-0.10}$ \\
    \hline
        \multirow{4}{*}{Wide}
        & \multirow{2}{*}{inner}  
        & 1.25 & $67.92^{+0.03}_{-0.03}$ & $0.710^{+0.008}_{-0.008}$ & $4.99^{+0.06}_{-0.06}$ & $4.62^{+0.10}_{-0.10}$ & $1.00^{+0.04}_{-0.04}$ & $1.14^{+0.04}_{-0.05}$ & \multirow{2}{*}{$1.01^{+0.02}_{-0.02}$} \\
        & & 2.14 & $67.88^{+0.05}_{-0.05}$ & $0.351^{+0.006}_{-0.006}$ & $4.93^{+0.09}_{-0.08}$ & $3.87^{+0.17}_{-0.17}$ & $1.27^{+0.06}_{-0.06}$ & $1.26^{+0.06}_{-0.06}$ \\
        & \multirow{2}{*}{outer}  
        & 1.25 & $100.77^{+0.05}_{-0.05}$ & $0.393^{+0.004}_{-0.004}$ & $5.46^{+0.07}_{-0.07}$ & $<0.85(0.42^{+0.43}_{-0.29})$ & $0.49^{+0.07}_{-0.07}$ & $1.21^{+0.07}_{-0.07}$ & \multirow{2}{*}{$1.25^{+0.03}_{-0.03}$}  \\
        & & 2.14 & $100.69^{+0.05}_{-0.05}$ & $0.292^{+0.006}_{-0.005}$ & $4.37^{+0.08}_{-0.10}$ & $<1.40(0.85^{+0.55}_{-0.57})$ & $0.82^{+0.07}_{-0.07}$ & $0.94^{+0.09}_{-0.09}$ \\
    \hline
        \multirow{4}{*}{Full angle}
        & \multirow{2}{*}{inner}  
          & 1.25 & $67.83^{+0.03}_{-0.03}$ & $0.635^{+0.007}_{-0.007}$ & $5.35^{+0.07}_{-0.07}$ & $3.40^{+0.14}_{-0.14}$ & $0.84^{+0.04}_{-0.05}$ & $1.13^{+0.04}_{-0.05}$ & \multirow{2}{*}{$1.03^{+0.02}_{-0.02}$} \\
        & & 2.14 & $67.77^{+0.05}_{-0.05}$ & $0.319^{+0.005}_{-0.005}$ & $5.18^{+0.10}_{-0.10}$ & $2.74^{+0.23}_{-0.25}$ & $1.14^{+0.06}_{-0.06}$ & $1.19^{+0.06}_{-0.06}$  \\
        & \multirow{2}{*}{outer}  
          & 1.25 & $100.50^{+0.04}_{-0.04}$ & $0.429^{+0.005}_{-0.004}$ & $4.57^{+0.07}_{-0.07}$ & $<0.63(0.30^{+0.31}_{-0.21})$ & $0.34^{+0.06}_{-0.06}$ & $1.18^{+0.06}_{-0.06}$ & \multirow{2}{*}{$1.23^{+0.03}_{-0.03}$}  \\
        & & 2.14 & $100.42^{+0.04}_{-0.04}$ & $0.318^{+0.005}_{-0.005}$ & $3.71^{+0.08}_{-0.08}$ & $<0.63(0.31^{+0.32}_{-0.22})$ & $0.84^{+0.07}_{-0.06}$ & $0.93^{+0.08}_{-0.08}$ \\
    \hline
        \multirow{4}{*}{High temp}
        & \multirow{2}{*}{inner}  
          & 1.25 & $67.57^{+0.04}_{-0.05}$ & $1.083^{+0.011}_{-0.011} \times 10^{-2}$ & $5.71^{+0.08}_{-0.08}$ & $3.62^{+0.12}_{-0.13}$ & $0.76^{+0.05}_{-0.05}$ & $1.15^{+0.06}_{-0.06}$ & \multirow{2}{*}{$1.14^{+0.03}_{-0.03}$} \\
        & & 2.14 & $67.60^{+0.07}_{-0.07}$ & $0.716^{+0.013}_{-0.012} \times 10^{-2}$ & $5.01^{+0.12}_{-0.11}$ & $3.51^{+0.19}_{-0.19}$ & $1.07^{+0.07}_{-0.07}$ & $1.26^{+0.07}_{-0.08}$ & \\
        & \multirow{2}{*}{outer}  
          & 1.25 & $100.26^{+0.06}_{-0.06}$ & $0.583^{+0.006}_{-0.006} \times 10^{-2}$ & $4.96^{+0.10}_{-0.09}$ & $<0.47(0.23^{+0.24}_{-0.16})$ & $0.48^{+0.08}_{-0.08}$ & $1.30^{+0.08}_{-0.08}$ & \multirow{2}{*}{$1.29^{+0.04}_{-0.04}$}  \\
        & & 2.14 & $100.36^{+0.06}_{-0.06}$ & $0.521^{+0.009}_{-0.008} \times 10^{-2}$ & $3.84^{+0.09}_{-0.10}$ & $<0.86(0.45^{+0.41}_{-0.31})$ & $0.86^{+0.09}_{-0.08}$ & $0.85^{+0.10}_{-0.10}$
    \enddata
    \tablecomments{
        Fitted parameters for the different fitting regions and the different temperature model.
        The median and 1$\sigma$ range of the MCMC results are shown.
        The label ``Fiducial'' is the fitting result with the fiducial setting discussed in the main text, ``Wide'' is the fitting result with the wider fitting region in the radial direction, ``Full angle'' is the fitting result with the fitting region including the non-axisymmetric structure in the southwest, and ``High temp'' is the fitting result with the temperature distribution as the high-temperature limit of 1000 K. 
    }
\end{deluxetable*}

\begin{deluxetable*}{ccccccccccc}
    \tablecaption{fitted parameters\label{tab:MLE}}
    \tablewidth{0pt}
    \tablehead{\colhead{model} & \colhead{ring} & \colhead{wavelength} & \colhead{$r_0$} & \colhead{$\tau_0$} & \colhead{$w_0$} & \colhead{$h_0$} & \colhead{$x_{\mathrm{cen}}$} & \colhead{$y_{\mathrm{cen}}$} & \colhead{$_\mathrm{red-}\chi^2$} \\
    \colhead{} & \colhead{} & \colhead{[mm]} & \colhead{[au]} & \colhead{} & \colhead{[au]} & \colhead{[au]} & \colhead{[au]} & \colhead{[au]}}
    \startdata
\multirow{4}{*}{Fiducial}
    & \multirow{2}{*}{inner}  
    & 1.25 &  $67.82$ & $0.718$ & $4.82$ & $4.28$ & $0.80$ & $1.19$ & $7.7$ \\
    & & 2.14 & $67.79$ & $0.364$ & $4.64$ & $3.76$ & $1.08$ & $1.27$ & $5.0$ \\
    & \multirow{2}{*}{outer}  
    & 1.25 & $100.41$ & $0.423$ & $4.67$ & $0.01$ & $0.49$ & $1.31$ & $2.7$ \\
    & & 2.14 & $100.45$ & $0.320$ & $3.70$ & $0.00$ & $0.85$ & $0.85$ & $4.5$ \\
\hline
    \multirow{4}{*}{Wide}
    & \multirow{2}{*}{inner}  
    & 1.25 & $67.92$ & $0.710$ & $4.99$ & $4.62$ & $1.00$ & $1.14$ & $6.5$ \\
    & & 2.14 & $67.89$ & $0.351$ & $4.93$ & $3.87$ & $1.26$ & $1.25$ & $4.3$ \\
    & \multirow{2}{*}{outer}  
    & 1.25 & $100.77$ & $0.392$ & $5.48$ & $0.03$ & $0.50$ & $1.21$ & $4.3$ \\
    & & 2.14 & $100.69$ & $0.293$ & $4.36$ & $0.98$ & $0.82$ & $0.95$ & $4.3$ \\
\hline
    \multirow{4}{*}{Full angle}
    & \multirow{2}{*}{inner}
    & 1.25 & $67.83$ & $0.635$ & $5.35$ & $3.40$ & $0.84$ & $1.13$ & $9.7$ \\
    & & 2.14 & $67.77$ & $0.320$ & $5.17$ & $2.75$ & $1.13$ & $1.19$ & $6.6$ \\
    & \multirow{2}{*}{outer}  
    & 1.25 & $100.50$ & $0.428$ & $4.58$ & $0.01$ & $0.34$ & $1.18$ & $2.4$ \\
    & & 2.14 & $100.42$ & $0.317$ & $3.71$ & $0.00$ & $0.84$ & $0.94$ & $3.8$ \\
\hline
    \multirow{4}{*}{High temp}
    & \multirow{2}{*}{inner}  
    & 1.25 & $67.57$ & $1.083 \times 10^{-2}$ & $5.71$ & $3.62$ & $0.76$ & $1.15$ & $8.5$ \\
    & & 2.14 & $67.60$ & $0.717 \times 10^{-2}$ & $5.00$ & $3.52$ & $1.07$ & $1.26$ & $5.0$ \\
    & \multirow{2}{*}{outer}  
    & 1.25 & $100.26$ & $0.583 \times 10^{-2}$ & $4.96$ & $0.00$ & $0.48$ & $1.30$ & $2.7$ \\
    & & 2.14 & $100.37$ & $0.518 \times 10^{-2}$ & $3.86$ & $0.00$ & $0.85$ & $0.85$ & $4.5$ \\ 
\enddata
\tablecomments{
    Maximum likelihood estimators, which maximaize the probability.
}
\end{deluxetable*}

\section{the case of different dust properties} \label{sec:porous}

We discuss the cases of different dust properties: different dust compositions and different dust porosities. 
Note that the following discussions are not intended to constrain the dust porosity or the dust composition but rather to constrain the dust size distribution given each dust property.

The porosity of the dust grains is not well constrained.
The high polarization fraction in millimeter observations suggests compact grains \citep[e.g.,][]{Tazaki2019}, while the dust collision simulations suggest that the dust grains grow into porous aggregates \citep[e.g.,][]{Kataoka2013}.
Since there is currently no consensus on the dust porosity, we assumed the compact dust grains in the main analysis, and we also provided a similar discussion assuming porous grains.

We perform the same model calculations as in Section \ref{sec:modeling} for porous dust grains with the filling factor $f=0.01$, i.e., porosity $p=0.99$.
Figure \ref{fig:dust_size_constraint_porous} shows the model width ratio and the opacity spectral index between Band 6 and Band 4.
In contrast to the compact grains case, there are no local minima of the ring width ratio with $w_{\mathrm{B6}}/w_{\mathrm{B4}}<1$ and local maximum with the opacity spectral index $\beta > 2$ when the maximum dust size is a few hundred micrometers.
This is because the porous grains do not have an opacity peak, as shown in Figure \ref{fig:opacity}.
By comparing with the observations, we constrained that $8\ \mathrm{mm} < a_{\mathrm{max}} < 4 \times 10^1 \ \mathrm{mm}$ and $p < 3.8$ the inner ring, and $a_{\mathrm{max}} > 2 \times 10^2 \ \mathrm{mm}$ and $3.5 < p < 3.7$ in the outer ring.

The dust compositions are also not well constrained.
In particular, the model of carbonaceous materials affects the absorption and scattering opacity, resulting in different albedos \citep[Figure 10 of ][]{Birnstiel2018}.
The refractory organics have low absorption opacity and high albedo \citep{Henning1996}, while the amorphous carbon or graphite have high absorption opacity and low albedo \citep{Zubko1996,Jager1998, Draine2006}.
The high polarization fraction in millimeter observations due to self-scattering suggests high albedo \citep{Kataoka2015,Kataoka2016b_HD}, while the weak intensity in infrared scattering observations suggests low albedo \citep{Tazaki2023}.
Thus, the dust composition and the dust opacity remain unknown.

We also perform the same model calculations as in Section \ref{sec:modeling} for the \citet{Ricci2010} dust model, which uses the \citet{Zubko1996} amorphous carbon model.
Figure \ref{fig:dust_size_constraint_porous} shows the modeling results and the constraints on the dust size distributions.
The ring width ratio and the opacity spectral index show similar trends to those for the DSHARP dust assumed in Section \ref{sec:dust_size}, while the overall trend of the figure is shifted towards the smaller dust grain side.
We constrained that $0.5\ \mathrm{mm} < a_{\mathrm{max}} < 0.7 \ \mathrm{mm}$ and $p < 2.3$ in the inner ring and $8 \ \mathrm{mm} < a_{\mathrm{max}} < 2 \times 10^2 \ \mathrm{mm}$ and $3.4 < p < 3.8$ in the outer ring.

Thus, the constrained parameter space depends on the dust model.
However, the exponent of the dust size distribution in the outer ring remains at $p \sim 3.5$, and the maximum dust size is consistently larger in the outer ring than in the inner ring for each dust model.
Therefore, we conclude that these findings are robust.

\section{Dependence of Comparing Region} \label{app:comparing}

Since the choice of the fitting region can affect the fitting results, we performed the fitting with different regions for comparison and to check the robustness of our result.
We performed the same fitting as in Section \ref{sec:obs} for different fitting regions: radially wider fitting region and azimuthally wider fitting region.
In the radially wider case, we performed the fitting for the radial range between 55 to 80 au in the inner ring and between 90 to 110 au in the outer ring excluding $90^{\circ}$ for the southeast direction.
In the azimuthally wider case, we performed the fitting in the same radial range as in Section \ref{sec:obs} for full azimuthal direction without excluding the southeast direction.

We show the results in table \ref{tab:fitted_app}.
When the observations are fitted for wider radial ranges, the fitted ring width tends to be larger since the observed intensity is greater at the edge of the ring, as shown in Figure \ref{fig:radial_2band}.
However, for any fitting, the dust ring width in the outer ring is consistently narrower in Band 4 than in Band 6, and the dust ring width in the inner ring remains comparable between the two bands.
Therefore, these results are robust.

\section{Dependence of Temperature Model} \label{app:temperature}

Since the temperature model assumed in the fitting can affect the fitting results, we perform the fitting using different temperature models to check the robustness of our main results.
Here, we assume that the disk temperature is $T=1000\ \mathrm{K}$ throughout the disk as a high-temperature limit.
We present the results in Table \ref{tab:fitted_app}.
In this model, we derive a smaller value for the ring distance due to the lack of a temperature gradient.
However, the dust ring width and width ratio remain consistent with the model in the main text. The ring width ratios in the main text are robust results.

\section{Probalility Distribution and Maximum Likelihood Estimator} \label{app:mcmc}

\begin{figure*}
    \begin{center}
        \includegraphics[width=17 cm]{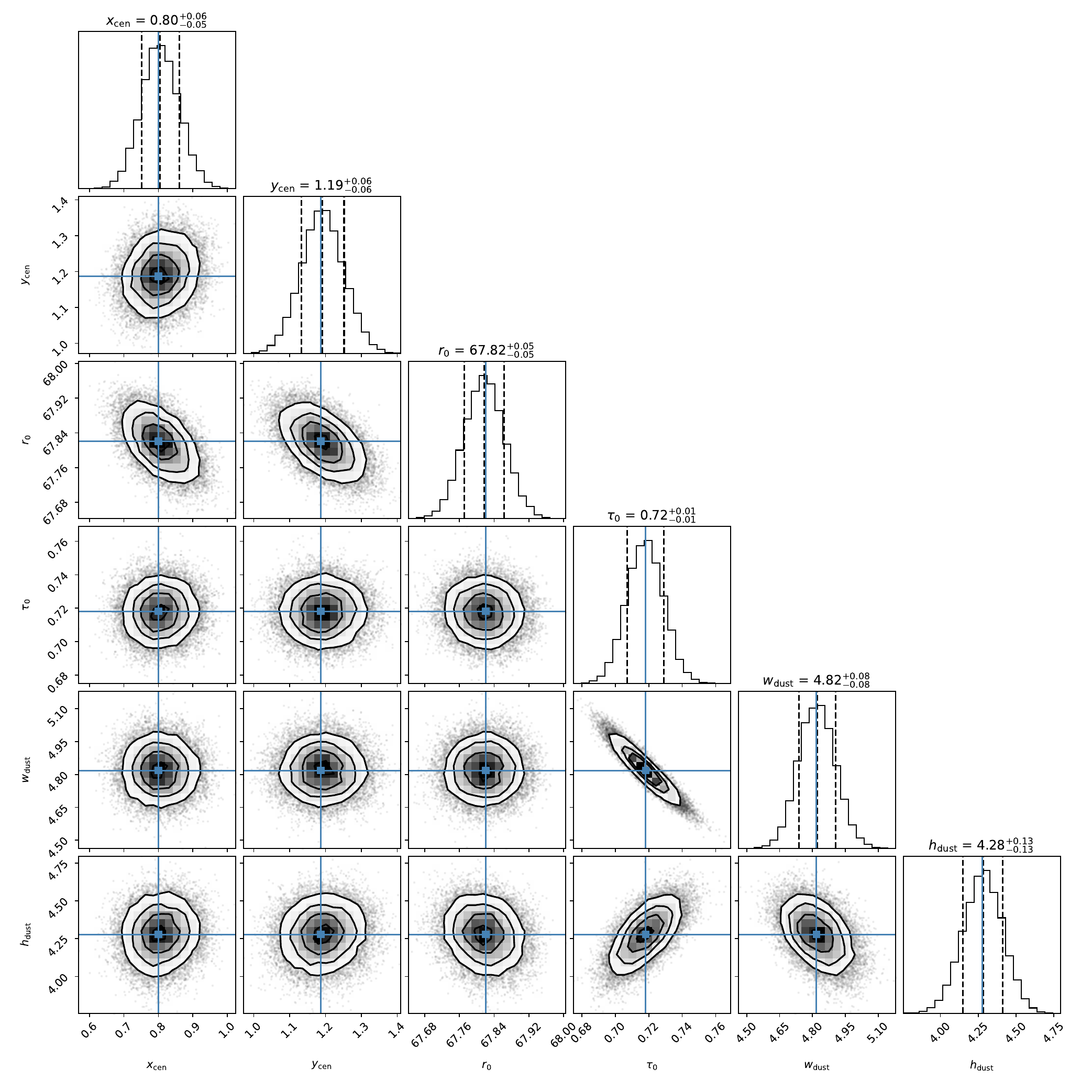}
        \caption{
            The corner plot for the inner ring in Band 6.
            The dotted lines in the top panels show the 16th, 50th, and 84th percentiles of the marginal distributions.
            The blue lines show the parameter set that maximizes the probability.
           \label{fig:corner1}}
    \end{center}
\end{figure*}

\begin{figure*}
    \begin{center}
        \includegraphics[width=17 cm]{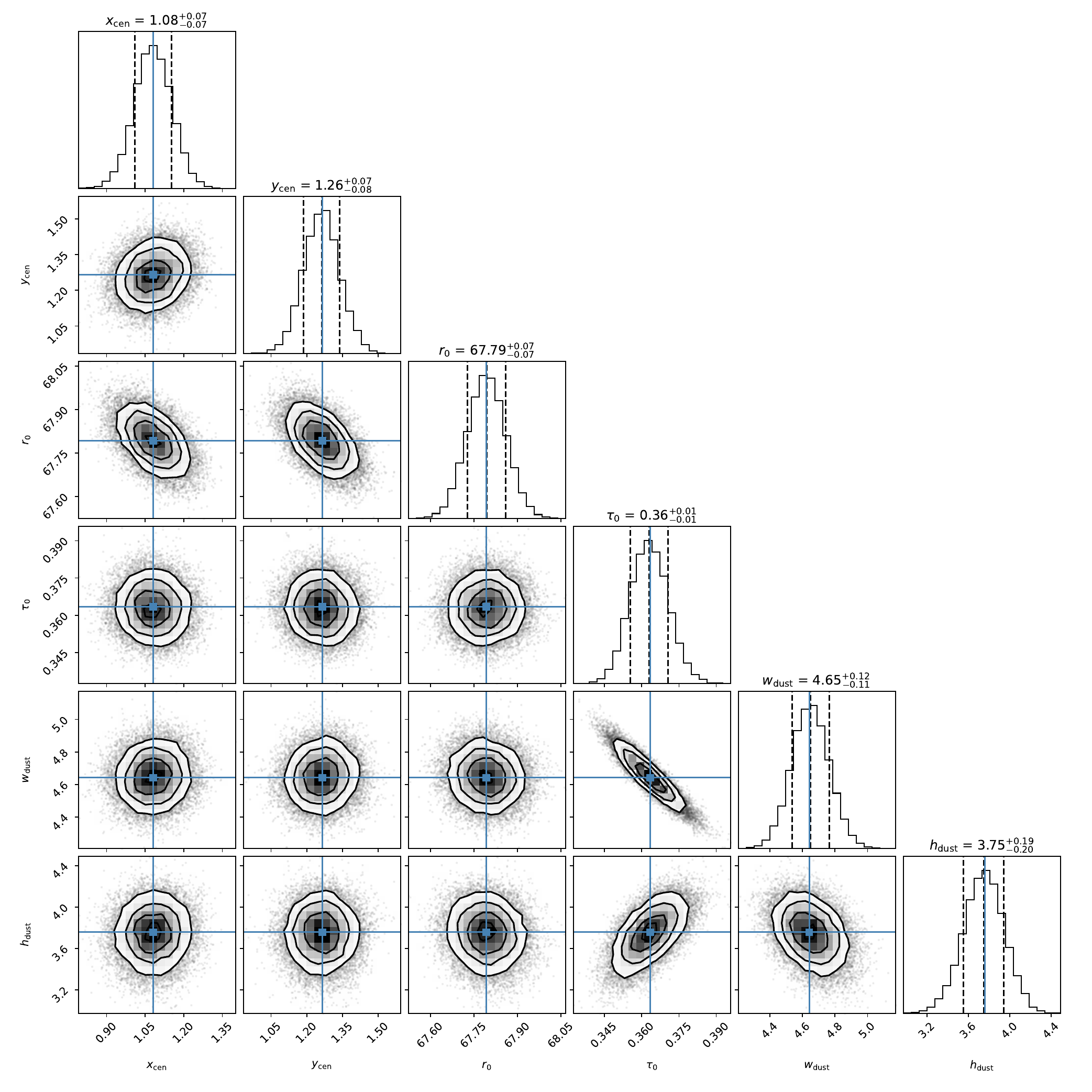}
        \caption{
            The same as Figure \ref{fig:corner1} but for the in ring in Band 4.
           \label{fig:corner2}}
    \end{center}
\end{figure*}

We show the corner plots of the MCMC sampling.
Figure \ref{fig:corner1} and \ref{fig:corner2} show the corner plot for the inner ring in Band 6 and Band 4, respectively.
The blue lines show the maximum likelihood estimator, which is the parameter set that maximizes the probability.
The MCMC sampling results are converged to Gaussian distributions.

\begin{figure*}
    \begin{center}
        \includegraphics[width=17 cm]{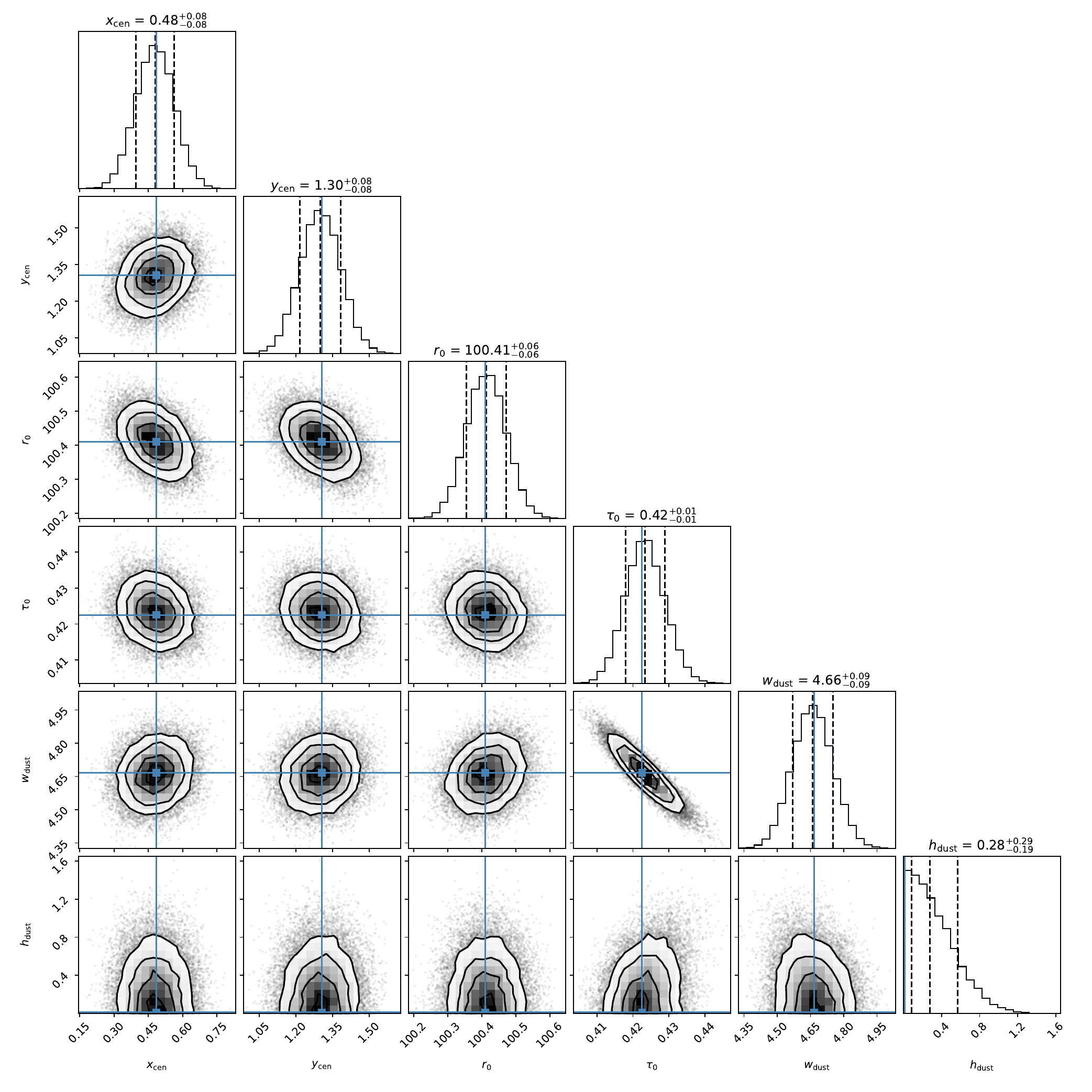}
        \caption{
            The same as Figure \ref{fig:corner1} but for the outer ring in Band 6.
           \label{fig:corner3}}
    \end{center}
\end{figure*}

\begin{figure*}
    \begin{center}
        \includegraphics[width=17 cm]{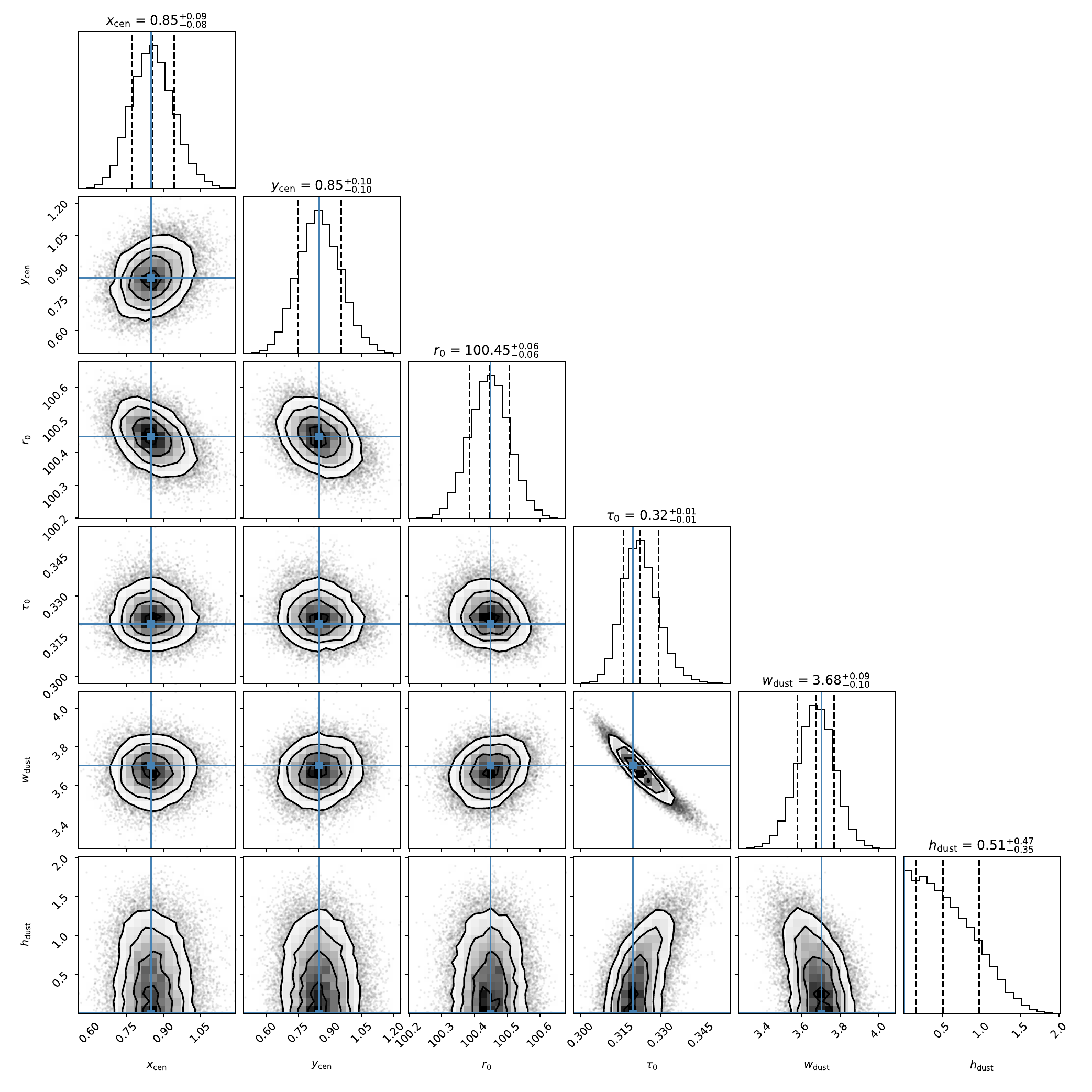}
        \caption{
            The same as Figure \ref{fig:corner1} but for the outer ring in Band 4.
           \label{fig:corner4}}
    \end{center}
\end{figure*}

\begin{figure*}
    \begin{center}
        \includegraphics[width=17 cm]{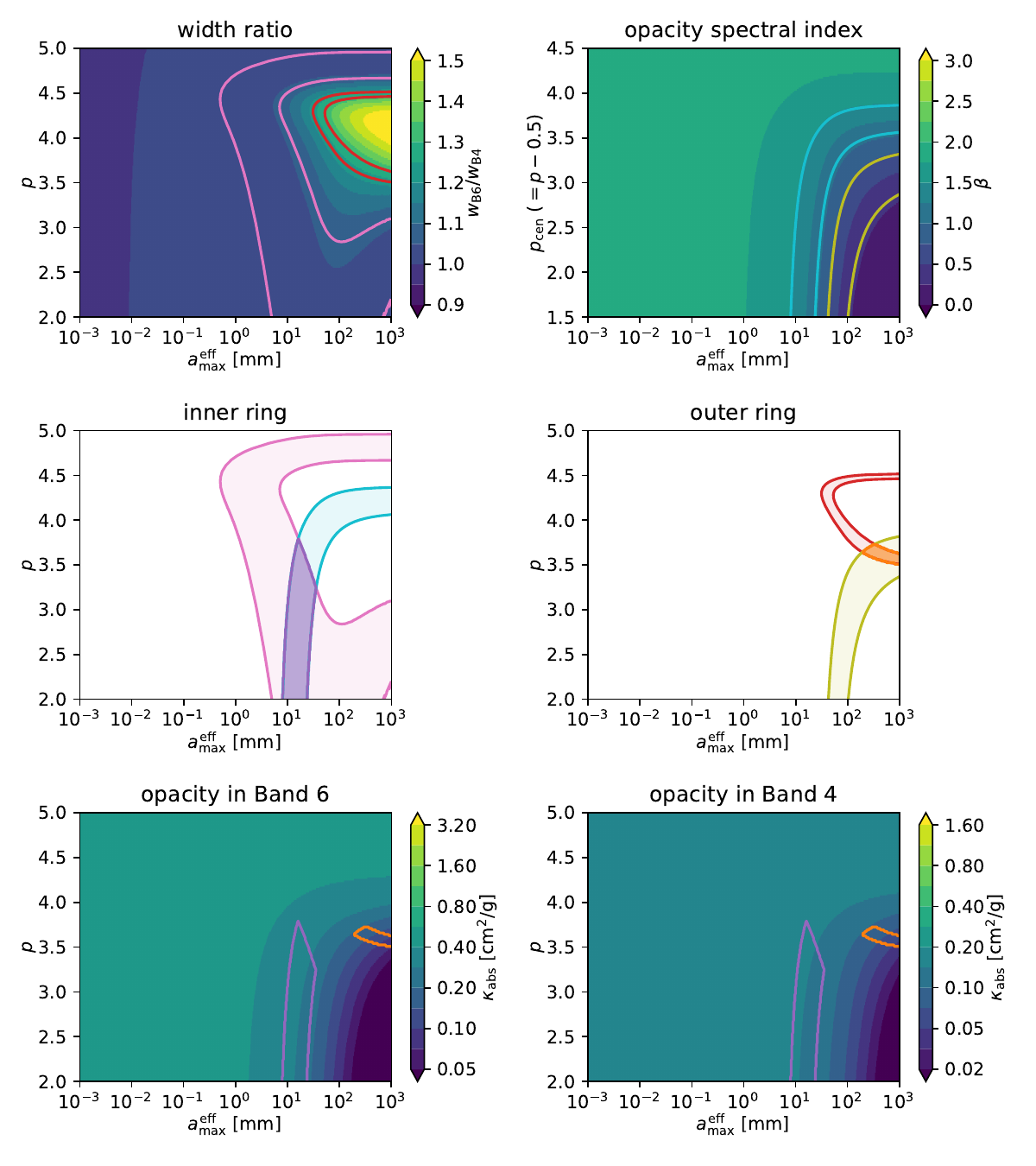}
        \caption{
            The same as Figure \ref{fig:width_ratio}, but assuming porous dust grains with the filling factor $f=0.01$.
           \label{fig:dust_size_constraint_porous}}
    \end{center}
\end{figure*}

Figure \ref{fig:corner3} and \ref{fig:corner4} show the corner plot for the outer ring in Band 6 and Band 4, respectively.
The maximum likelihood estimator of the dust scale height is close to zero, and it is out of the 1 sigma error range of the marginal distribution.
In this case, we use the upper limit of the 1 sigma error range of the marginal distribution as the upper limit of the fitting.

We show the maximum likelihood estimator in Table \ref{tab:MLE}.
By comparing with Table \ref{tab:fitted_app}, the maximum likelihood estimator is similar to the 50th percentile of the marginal distribution for all parameters except for the dust scale height in the outer ring.
The maximum likelihood estimators of the dust scale height is close to zero in the outer ring in all models, and fall outside the 1 sigma error range of the marginal distribution. Consequently, we utilize only the upper limit as a reliable constraint for the dust scale height.

\begin{figure*}
    \begin{center}
        \includegraphics[width=17 cm]{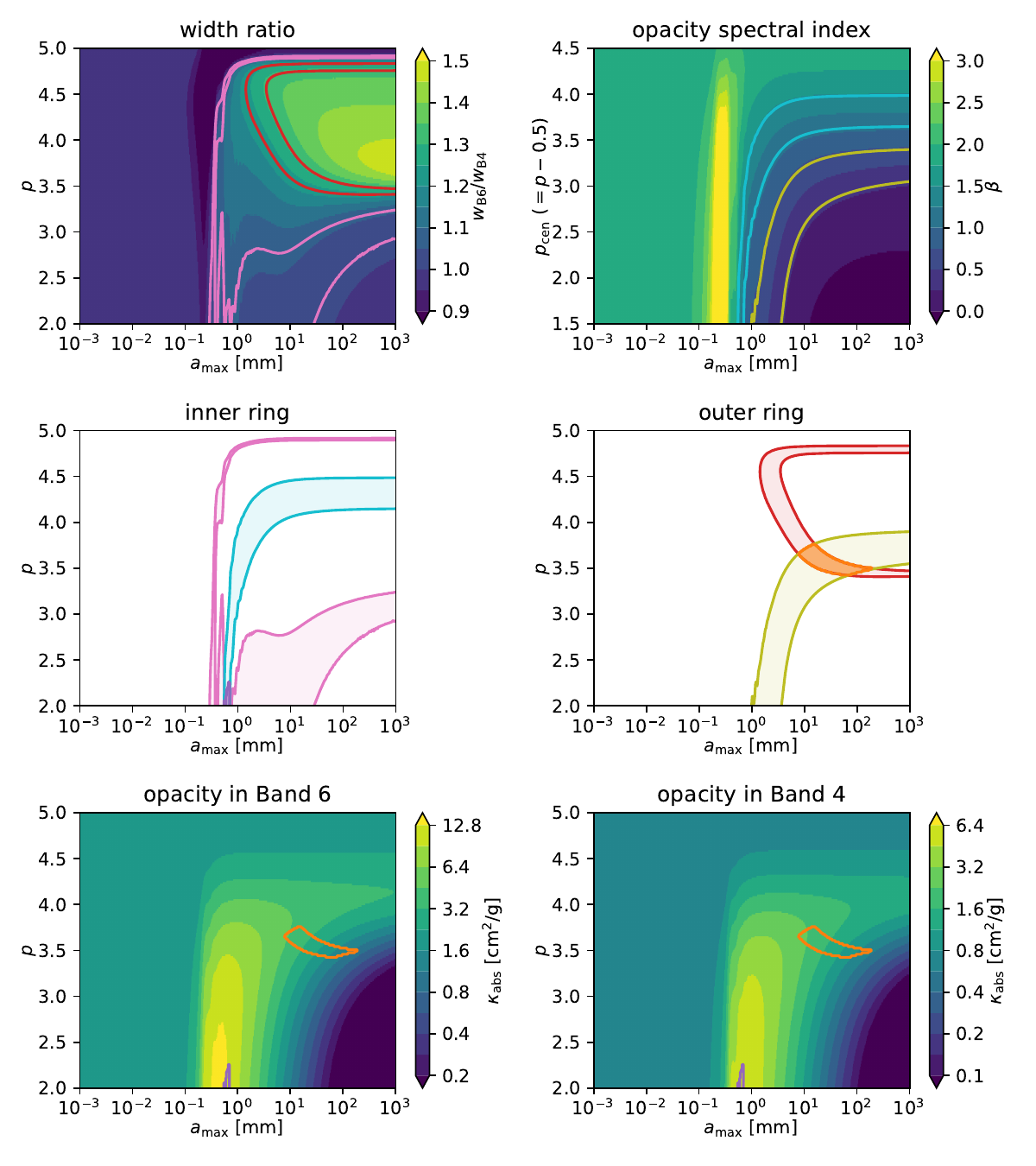}
        \caption{
            The same as Figure \ref{fig:width_ratio}, but assuming dust model by \citet{Ricci2010}, which assumes amorphous carbon as a carbonaceous material.
           \label{fig:dust_size_constraint_ricci}}
    \end{center}
\end{figure*}


\bibliography{doi_citation}{}
\bibliographystyle{aasjournal}

\end{CJK*}
\end{document}